%% file: main.tex
\documentclass[preprint,12pt]{elsarticle}

\usepackage{amsmath}
\usepackage{amssymb}
\usepackage{amsthm}
\usepackage{bm}
\usepackage[Euler]{upgreek}
\usepackage{graphicx}
\usepackage{booktabs}
\usepackage{quantikz}
\usepackage[colorlinks=true,linkcolor=blue,citecolor=blue,urlcolor=blue]{hyperref}
\usepackage{microtype}
\usepackage{orcidlink}
\usepackage{setspace}

\graphicspath{{figures/}}
\theoremstyle{plain}
\newtheorem{lemma}{Lemma}
\newtheorem{proposition}{Proposition}
\newtheorem{theorem}{Theorem}
\theoremstyle{definition}
\newtheorem{remark}{Remark}
\newtheorem{example}{Example}

\newcommand{\diag}{\operatorname{diag}}
\newcommand{\lamax}{\lambda_{\max}}
\newcommand{\sbar}{\bar\sigma}
\newcommand{\nL}{n_{\mathrm{L}}}
\newcommand{\QFT}{\mathrm{QFT}}
\providecommand{\ket}[1]{\lvert #1\rangle}
\providecommand{\bra}[1]{\langle #1\rvert}
\newcommand{\Df}{\mathbf{D}^{\mathrm f}}
\newcommand{\Db}{\mathbf{D}^{\mathrm b}}

\journal{Journal of Computational Physics}

\begin{document}

\begin{frontmatter}

\title{Quantum simulation of dissipative acoustic and elastic waves by
Schr\"odingerization: absorbing boundaries and explicit polynomial-size
circuits}

\author[esp]{Hoang Anh Nguyen\,\orcidlink{0000-0002-7504-5284}\corref{cor1}}
\ead{hoanganh\_nguyen@mines.edu}
\author[mse]{Ali Tura}
\cortext[cor1]{Corresponding author}
\address[esp]{School of Earth, Space and Planets, Colorado School of Mines,
Golden, Colorado 80401, USA}
\address[mse]{School of Minerals and Subsurface Energy, Colorado School of Mines,
Golden, Colorado 80401, USA}

\begin{abstract}
Explicit quantum circuits for the acoustic and elastic wave equations,
obtained by Hamiltonian simulation of their first-order forms, conserve
energy and therefore describe closed domains: an outgoing wave reflects
off the edge of the computational region or, with periodic boundaries,
wraps around and re-enters from the opposite side.  We open the
domain with the simplest physical absorber, a graded sponge layer in
which every wave field is damped at a rate that grows towards the outer
edge, and implement the resulting dissipative dynamics on a quantum
simulator by Schr\"odingerization, which turns a non-unitary linear
evolution into a family of unitary ones by means of a single auxiliary
variable.  We derive the whole chain from the partial differential
equation to the quantum circuit.  An independent finite-difference solver confirms that damping
all wave fields absorbs far better than the velocity-only
damping mask.  The damping step
factorizes exactly into polynomially many multi-controlled phase
gates.  We demonstrate that gate-level simulations on up to 25 qubits
agree with the classical solutions of the damped wave equations.
\end{abstract}


\end{frontmatter}

\input{sections/intro}
\input{sections/waves}
\input{sections/sponge}
\input{sections/schro}
\input{sections/circuits}
\input{sections/numerics}
\input{sections/discussion}

\section*{Code and data availability}
All code, data and scripts that produce the reported results are
available at
\url{https://github.com/x-repos/schodingerWave}.

\section*{CRediT authorship contribution statement}
\textbf{Hoang Anh Nguyen:} Conceptualization, Methodology, Software,
Formal analysis, Investigation, Visualization, Writing -- original
draft.  \textbf{Ali Tura:} Supervision, Writing -- review \& editing,
Funding acquisition.

\section*{Declaration of competing interest}
The authors declare that they have no known competing financial
interests or personal relationships that could have appeared to
influence the work reported in this paper.

\section*{Declaration of generative AI and AI-assisted technologies in the writing process}
During the preparation of this work the authors used Claude (Anthropic)
to assist with writing and editing the manuscript.  After using this
tool, the authors reviewed and edited the content as needed and take full
responsibility for the content of the published article.

\section*{Acknowledgements}
We would like to thank the sponsors of the Reservoir Characterization
Project (RCP) at the Colorado School of Mines for the financial support
of this work.

\appendix
\input{sections/appA}
\input{sections/appB}
\input{sections/appC}

\bibliographystyle{elsarticle-num-names}
\bibliography{refs}

\end{document}

%% file: sections/intro.tex
\section{Introduction}
\label{sec:intro}

Quantum algorithms for partial differential equations (PDEs) follow two
broad routes.  Linear-system and ordinary differential equation (ODE)
solvers treat a semi-discrete PDE
as a large linear problem~\cite{harrow2009linear,berry2017linear,childs2020spectral}; Hamiltonian simulation, where it applies,
evolves the discretized field directly as the state of a quantum
register~\cite{lloyd1996universal,low2019qubitization} and is the most economical route for time-dependent problems.
For waves it applies naturally: written in first-order form with
suitably scaled variables, the acoustic and the elastic wave equations
conserve an energy that is a Euclidean norm, so their semi-discrete
generators, the matrices that drive the time evolution, are
anti-Hermitian, $\partial_t\boldsymbol{\uppsi}=-i\mathbf{H}\boldsymbol{\uppsi}$ with
$\mathbf{H}=\mathbf{H}^{\dagger}$, in which $\boldsymbol{\uppsi}$ is the
vector of the discretized wave fields and $\mathbf{H}$ the Hamiltonian
matrix~\cite{costa2019wave,suau2021practical,schade2024elastic}.  Sato \emph{et
al.}~\cite{sato2024hamiltonian} turned this observation into explicit
circuits: the finite-difference operators decompose into a number of
ladder-operator strings that is linear in the number of qubits, each
string is exponentiated by a Bell-basis change and a single
multi-controlled rotation, and the Trotterized~\cite{trotter1959} evolution of the wave
equation costs a polynomial number of elementary gates per step.

Energy conservation is also a limitation.  A unitary evolution on a
finite grid cannot let energy leave the computational region.  With
reflecting boundaries (e.g.\ Dirichlet or Neumann) every outgoing wave
reflects off the edge; with periodic boundaries, the common choice in
quantum PDE algorithms because the difference operators then become
circulant and are diagonalized by the (quantum) Fourier
transform~\cite{costa2019wave,jin2025elastic}, it leaves through one
side and re-enters through the opposite one.  Either way it returns to
the area of interest.
Classical wave simulation in acoustics and electromagnetics
removes these reflections either by local absorbing boundary
conditions~\cite{clayton1977absorbing,givoli1991nonreflecting}
or by surrounding the region with an
absorbing layer, and the oldest and simplest such layer is the
sponge~\cite{israeli1981,kosloff1986}.  The same construction
appears in quantum dynamics as a complex absorbing
potential.  Damping is dissipation, and the evolution
is no longer unitary.

Schr\"odingerization~\cite{jin2024prl,jin2023schrodingerisation} is a
general device for simulating non-unitary linear dynamics,
$\dot{\mathbf{z}}=\mathbf{A}\mathbf{z}$, by unitary evolution; the linear combination of
Hamiltonian simulation~\cite{an2023lchs} is an alternative, and analog
approaches with bosonic modes have been proposed for nonlinear
dissipative equations~\cite{gan2025bosonic}.  Here
$\mathbf{z}(t)$ is the state vector, the dot denotes the time derivative,
and $\mathbf{A}$ is a general (non-Hermitian) matrix, so the evolution
$e^{\mathbf{A}t}$ is not unitary and cannot be run directly on a quantum
computer.  One adds a single auxiliary real variable $p$, applies the
warped phase transformation $\mathbf{w}(t,p)=e^{-p}\mathbf{z}(t)$, and
Fourier transforms in $p$; the result is a family of Hamiltonian systems
of Schr\"odinger form, one per Fourier mode, which a
quantum computer simulates in parallel on an extra register of a few
qubits that holds the discretized $p$ variable.  The solution is read
off by measuring that register.  The method has been applied to the heat, convection,
Fokker--Planck and Black--Scholes
equations~\cite{jin2023schrodingerisation}, to quantum dynamics with artificial
boundary conditions~\cite{jin2024abc}, to elastic waves on periodic
domains~\cite{jin2025elastic}, and has been compiled into explicit
circuits for several PDEs~\cite{hu2024quantum,jin2024heat}.  The
solution at the final time $T$ is recovered from the values of
$\mathbf{w}(T,p)$ at a chosen point $p=p^{*}$, and this recovery is valid
only if $p^{*}\ge\lamax^{+}(\mathbf{H}_1)\,T$~\cite{jin2024recovery}.
Here $\mathbf{H}_1=(\mathbf{A}+\mathbf{A}^{\dagger})/2$ is the Hermitian
part of $\mathbf{A}$ and $\lamax^{+}(\mathbf{H}_1)$ is its largest
eigenvalue, or zero if all eigenvalues are negative; a positive
eigenvalue therefore forces a large $p^{*}$ and makes the recovery
expensive.

In this paper we combine three ingredients, explicit wave circuits,
the sponge layer, and Schr\"odingerization, for the acoustic and the elastic wave equations.  The
sponge turns out to be the case in which every step of the
Schr\"odingerization pipeline is clean and can be analyzed sharply: its
Hermitian part is $\mathbf{H}_1=-\boldsymbol{\Sigma}$, where the diagonal
matrix $\boldsymbol{\Sigma}$ holds the local damping rates, so
$\mathbf{H}_1$ is diagonal and negative semidefinite, the recovery
threshold vanishes, the auxiliary dynamics is a pure
left-transport with speeds equal to the local damping rates, and the
dissipative factor of every Trotter step is a diagonal unitary.  A companion
paper~\cite{nguyen2026companion} gives explicit circuits for the acoustic
wave equation with a convolutional perfectly matched layer and uses the
sponge as a reference absorber, with the damping step compiled as a
generic diagonal gate.  The exact analysis of the Schr\"odingerized
sponge, the polynomial-size damping circuit and the elastic case are new
here.

The paper makes four contributions. First, we derive the path from
the PDE to a dissipative Hamiltonian system
(Secs.~\ref{sec:waves}--\ref{sec:sponge}): the energy-symmetrized
first-order forms of the acoustic and the isotropic elastic equations in
one, two and three dimensions, including the closed-form material square
root for the elastic case, the staggered forward/backward
finite-difference pairing and the boundary conditions that the truncated
difference operators imply, and the sponge generator
$\mathbf{A}=-i\mathbf{H}-\boldsymbol{\Sigma}$ of the damped system
$\dot{\boldsymbol{\uppsi}}=\mathbf{A}\boldsymbol{\uppsi}$.  We show that
the sponge is reflectionless for any profile in the one-dimensional
continuum and dissipative in every dimension
(Lemma~\ref{lem:dissipative}), and we verify with an independent
classical solver that damping all wave fields absorbs far better than
the common velocity-only mask.

Second, we show that the Schr\"odingerization of the sponge can be
analyzed in closed form (Sec.~\ref{sec:schro}).  For continuous $p$, recovery
is exact on all of $p\ge0$ (Proposition~\ref{prop:exact}).  The $p$
dynamics has a finite domain of dependence, so truncating the $p$ axis
periodically at $|p|<p_{\max}$ is exact for $p_{\max}\ge p^{*}+\sbar T$,
where $\sbar$ is the largest damping rate (Proposition~\ref{prop:dod}).
The success probability is
$\kappa_g\,\|\boldsymbol{\uppsi}(T)\|^{2}/\|\boldsymbol{\uppsi}_0\|^{2}$
in closed form, with a constant $\kappa_g\approx0.4$ set by the initial
profile in $p$ (Proposition~\ref{prop:prob}), so that the only loss
beyond this constant is the energy the sponge has absorbed.
Post-selecting one or two qubits of the $p$ register leaves the system
register in $\boldsymbol{\uppsi}(T)/\|\boldsymbol{\uppsi}(T)\|$.

Third, we construct a polynomial-size damping circuit
(Sec.~\ref{sec:structured}).  The damping factor of each time step
$\tau$, $e^{i\tau \mathbf{D}_\eta\otimes\boldsymbol{\Sigma}}$, where
$\mathbf{D}_\eta$ is the diagonal matrix of Fourier frequencies of the
$p$ variable, factorizes exactly into commuting multi-controlled phase
gates whose number is polynomial in the number of qubits $n$
(Theorem~\ref{thm:structured}).  It replaces a generic diagonal unitary
that costs $2^{n}-2$ controlled-NOT (CNOT) gates, so that the complete
Trotter step, wave part and damping part, is polynomial.

Fourth, we carry out end-to-end gate-level simulations
(Sec.~\ref{sec:numerics}) of absorbed acoustic and elastic waves on
17--25 qubits, including runs whose time loop contains only elementary
gates and multi-controlled phases.  They reproduce the classical
solutions of the damped wave equations with relative errors of
$5\times10^{-4}$--$1.7\times10^{-2}$ in the energy norm.

%% file: sections/waves.tex
\section{Wave equations as Hamiltonian systems}
\label{sec:waves}

This section rewrites the acoustic and the elastic wave equations as
$i\,\partial_t\boldsymbol{\uppsi}=\mathbf{H}\boldsymbol{\uppsi}$ with Hermitian $\mathbf{H}$, first in the continuum and
then on a grid that a register of qubits can hold.  Nothing here is
dissipative.

\subsection{Acoustic waves}
\label{sec:acoustic}

Consider the scalar wave equation in $d=1$, $2$ or $3$ dimensions with a
speed $c$, which we take constant for simplicity,
\begin{equation}
  \partial_t^{2}u = c^{2}\Delta u .
  \label{eq:wave}
\end{equation}
It is second order in time and not of Schr\"odinger form.  Introduce the
velocity $v=\partial_t u$ and the scaled gradient
$\mathbf{q}=(q_1,\dots,q_d)$ with $q_a = i c\,\partial_a u$.  Then
\begin{equation}
  i\,\partial_t v = c\sum_{a=1}^{d}\partial_a q_a,
  \qquad
  i\,\partial_t q_a = -c\,\partial_a v ,
  \label{eq:firstorder}
\end{equation}
since $i\partial_t v = ic^{2}\Delta u = c\sum_a\partial_a(ic\partial_a u)$ and
$i\partial_t q_a = i\cdot ic\,\partial_a\partial_t u = -c\,\partial_a v$.
With $\boldsymbol{\uppsi}=[v;\,q_1;\,\dots;\,q_d]$ this is
\begin{equation}
  i\,\partial_t\boldsymbol{\uppsi} = \mathbf{H}\boldsymbol{\uppsi},\qquad
  \mathbf{H} = \begin{bmatrix}
        0 & c\partial_1 & \cdots & c\partial_d\\
        -c\partial_1 & 0 & \cdots & 0\\
        \vdots & & \ddots & \\
        -c\partial_d & 0 & \cdots & 0
      \end{bmatrix}.
  \label{eq:Hcont}
\end{equation}
Because $\partial_a^{\dagger}=-\partial_a$ in $L^{2}$ (for decaying or
suitably closed boundary data), the lower-left blocks are the adjoints
of the upper-right ones and $\mathbf{H}$ is Hermitian.  The conserved quantity is
the acoustic energy
$\|\boldsymbol{\uppsi}\|^{2} = \|\partial_t u\|^{2}+c^{2}\|\nabla u\|^{2}$.  Hermiticity
comes from this energy conservation, not from the factor $i$: in the
real fields $v$ and $\mathbf{r}=c\nabla u$ the generator is already real
antisymmetric, so that $\mathbf{H}$ is Hermitian.  Writing
$\mathbf{q}=i\mathbf{r}$ is a unitary change of variables that only makes
$\mathbf{H}$ real; a dissipative generator, by contrast, cannot be made
Hermitian by inserting factors of $i$~\cite{jin2023schrodingerisation}.

Let each axis carry
$N=2^{n}$ nodes $x_j=jh$, $j=0,\dots,N-1$.  The forward and backward
differences are
\begin{equation}
  (\Df \mathbf{u})_j = \frac{u_{j+1}-u_j}{h},\qquad
  (\Db \mathbf{u})_j = \frac{u_j-u_{j-1}}{h},
  \label{eq:diffs}
\end{equation}
with the ghost values $u_N=u_{-1}=0$ that the finite register imposes.
As matrices, $\Df=(\mathbf{S}-\mathbf{I})/h$ and $\Db=(\mathbf{I}-\mathbf{S}^{\dagger})/h$, where $\mathbf{S}$ is the
non-periodic shift $\mathbf{S}\ket{j+1}=\ket{j}$, so that
\begin{equation}
  (\Df)^{\dagger} = -\Db
  \label{eq:adjoint}
\end{equation}
holds exactly, not just up to boundary terms.  This identity is
the discrete version of $\partial^{\dagger}=-\partial$ and is all that is
needed for a Hermitian discrete $\mathbf{H}$.

Put $v$ on the nodes and $q_a$ on the
half-cells $x_{j-1/2}$ of axis $a$, and discretize
$\mathbf{q}_a = ic\,\Db_a\mathbf{u}$ (a backward difference: it lives between $x_{j-1}$
and $x_j$).  Replacing $c\partial_a$ by $c\Df_a$ in the first row of
Eq.~\eqref{eq:Hcont} forces, by Eq.~\eqref{eq:adjoint}, the backward
difference $-c\partial_a\to(c\Df_a)^{\dagger}=-c\Db_a$ in the adjoint
rows, so
\begin{equation}
  \mathbf{H} = \sum_{a=1}^{d}\Bigl(\ket{0}\!\bra{a}\otimes c\Df_a
      + \ket{a}\!\bra{0}\otimes (c\Df_a)^{\dagger}\Bigr),
  \label{eq:Hdisc}
\end{equation}
where $\ket{b}\!\bra{b'}$ acts on the block label ($b=0$ for $v$,
$b=a$ for $q_a$) and $\Df_a$ is $\Df$ on axis $a$ tensored with
identities.  Eliminating $\mathbf{q}$ gives
$\partial_t^{2}\mathbf{u} = c^{2}\sum_a\Df_a\Db_a\mathbf{u}$, and $\Df\Db$ is the
standard three-point Laplacian $(u_{j+1}-2u_j+u_{j-1})/h^{2}$; the
staggered pair therefore reproduces the classical second-order scheme.

The truncation of the shift at the ends of
the register is a boundary condition.  The first row of $\Df\Db$ reads
$(-2u_0+u_1)/h^{2}$: a Dirichlet condition $u_{-1}=0$ one node outside
the left edge.  The last row reads $(u_{N-2}-u_{N-1})/h^{2}$: the flux
$(\Db\mathbf{u})_N$ is set to zero, a Neumann condition at $x_{N-1/2}$.  These
are the mixed conditions of \citet{sato2024hamiltonian}.  Either way
the edge is a perfect reflector: no energy leaves.  Closing the shift
periodically instead (adding the wrap-around term $\ket{N-1}\!\bra{0}$
to $\mathbf{S}$) also gives a Hermitian $\mathbf{H}$, but then a wave
leaving through one edge re-enters through the opposite one; again no
energy leaves.  This is the problem the sponge solves.  We do not use central differences: on a
non-periodic grid the central difference decouples even and odd nodes, whereas the forward/backward pair couples
them.

The fields are stacked in blocks, one per field, and
the number of blocks must be a power of two, so it is padded with zero
blocks where needed.  In $d=1$ the state
$\boldsymbol{\uppsi}=[\mathbf{v};\,\mathbf{q}]$ has two blocks of $N_x$
values and occupies $n_x+1$ qubits.  In $d=2$ the three fields are
padded to four blocks,
$\boldsymbol{\uppsi}=[\mathbf{v};\,\mathbf{q}_x;\,\mathbf{q}_y;\,\mathbf{0}]$,
of $N_xN_y$ values each, on $n_x+n_y+2$ qubits; the two block-label
qubits are the most significant and the $x$ index is more significant
than $y$.  In $d=3$ the four fields
$[\mathbf{v};\,\mathbf{q}_x;\,\mathbf{q}_y;\,\mathbf{q}_z]$ fill the two
label qubits exactly, without padding, and $\boldsymbol{\uppsi}$ occupies
$n_x+n_y+n_z+2$ qubits.

\subsection{Elastic waves}
\label{sec:elastic}

Isotropic elastodynamics in velocity--stress form reads
\begin{equation}
  \rho\,\partial_t \mathbf{v} = \nabla\!\cdot\!\boldsymbol{\upsigma},
  \qquad
  \partial_t\boldsymbol{\upsigma} = \mathbf{C}:\boldsymbol{\upvarepsilon}(\mathbf{v}),
  \label{eq:elastic}
\end{equation}
with particle velocity $\mathbf{v}$, symmetric stress $\boldsymbol{\upsigma}$, strain rate
$\boldsymbol{\upvarepsilon}(\mathbf{v})=(\nabla\mathbf{v}+\nabla\mathbf{v}^{\mathsf T})/2$, density $\rho$ and
isotropic stiffness $\mathbf{C}$ with Lam\'e parameters $\lambda,\mu$.  The P and
S speeds are $c_p=\sqrt{(\lambda+2\mu)/\rho}$ and $c_s=\sqrt{\mu/\rho}$.
This is the form in which seismic wavefields are computed
classically~\cite{virieux1986psv,graves1996,moczo2007finite}.

We present the two-dimensional case, which is the one we simulate, in
detail; one and three dimensions follow at the end of this subsection.

Collect the stress in the $\sqrt2$-weighted Voigt
vector; in two dimensions
$\tilde{\boldsymbol{\upsigma}}=(\sigma_{xx},\sigma_{yy},\sqrt2\,\sigma_{xy})^{\mathsf T}$.
Then $\partial_t\tilde{\boldsymbol{\upsigma}}=\tilde{\mathbf{C}} \mathbf{E}\mathbf{v}$ with
\begin{equation}
  \tilde{\mathbf{C}}=\begin{bmatrix}
    \lambda+2\mu & \lambda & 0\\
    \lambda & \lambda+2\mu & 0\\
    0 & 0 & 2\mu\end{bmatrix},
  \qquad
  \mathbf{E}=\begin{bmatrix}
    \partial_x & 0\\ 0 & \partial_y\\
    \partial_y/\sqrt2 & \partial_x/\sqrt2\end{bmatrix},
  \label{eq:voigt}
\end{equation}
and $\nabla\!\cdot\!\boldsymbol{\upsigma} = \mathbf{E}^{\mathsf T}\tilde{\boldsymbol{\upsigma}}$, where
$\mathbf{E}^{\mathsf T}$ is the transpose of the operator matrix (not its
adjoint).  The $\sqrt2$ weight is what makes the shear row of $\mathbf{E}$ and the
divergence share one operator: $\mathbf{E}^{\mathsf T}=-\mathbf{E}^{\dagger}$.

Scale $\tilde{\mathbf{v}}=\sqrt\rho\,\mathbf{v}$ and
$\mathbf{s}=\tilde{\mathbf{C}}^{-1/2}\tilde{\boldsymbol{\upsigma}}$, where $\tilde{\mathbf{C}}^{1/2}$ is the symmetric
positive square root.  Then
\begin{equation}
  \partial_t\mathbf{s} = \mathbf{G}\tilde{\mathbf{v}},\qquad
  \partial_t\tilde{\mathbf{v}} = -\mathbf{G}^{\dagger}\mathbf{s},\qquad
  \mathbf{G}=\frac{1}{\sqrt\rho}\,\tilde{\mathbf{C}}^{1/2}\mathbf{E},
  \label{eq:elasticpair}
\end{equation}
because $\partial_t\mathbf{s}=\tilde{\mathbf{C}}^{-1/2}\tilde{\mathbf{C}} \mathbf{E}\mathbf{v}=\tilde{\mathbf{C}}^{1/2}\mathbf{E}\tilde{\mathbf{v}}
/\sqrt\rho$ and
$\partial_t\tilde{\mathbf{v}} = \mathbf{E}^{\mathsf T}\tilde{\boldsymbol{\upsigma}}/\sqrt\rho
=-\mathbf{E}^{\dagger}\tilde{\mathbf{C}}^{1/2}\mathbf{s}/\sqrt\rho=-\mathbf{G}^{\dagger}\mathbf{s}$.  The energy
$\tfrac12(\rho|\mathbf{v}|^{2}+\boldsymbol{\upsigma}:\mathbf{C}^{-1}:\boldsymbol{\upsigma})$ becomes
$\tfrac12(|\tilde{\mathbf{v}}|^{2}+|\mathbf{s}|^{2})$, a Euclidean norm, and with
$\boldsymbol{\uppsi}=[\tilde{\mathbf{v}};\,\mathbf{s}]$
\begin{equation}
  i\,\partial_t\boldsymbol{\uppsi} = \mathbf{H}\boldsymbol{\uppsi},\qquad
  \mathbf{H} = i\begin{bmatrix} 0 & -\mathbf{G}^{\dagger}\\ \mathbf{G} & 0\end{bmatrix}
  \label{eq:Helastic}
\end{equation}
is Hermitian.  The construction is the elastic analogue of
Eq.~\eqref{eq:Hcont} and follows \citet{jin2025elastic} up to a
constant change of variables.

The normal block of $\tilde{\mathbf{C}}$
is $\lambda\mathbf 1\mathbf 1^{\mathsf T}+2\mu\mathbf{I}$, whose eigenvectors
are $\mathbf 1$ and its orthogonal complement.  Hence, with
$\boldsymbol{\Pi}=\mathbf 1\mathbf 1^{\mathsf T}/k$ the projector onto $\mathbf 1$
in the $k$ normal components ($k=2$ in two, $k=3$ in three
dimensions),
\begin{equation}
  \tilde{\mathbf{C}}^{1/2} =
  \begin{bmatrix}
    \sqrt{k\lambda+2\mu}\;\boldsymbol{\Pi}+\sqrt{2\mu}\,(\mathbf{I}-\boldsymbol{\Pi}) & 0\\
    0 & \sqrt{2\mu}\,\mathbf{I}
  \end{bmatrix}.
  \label{eq:csqrt}
\end{equation}
In two dimensions this is the matrix with diagonal $(s_1+s_2)/2$,
off-diagonal $(s_1-s_2)/2$ in the normal block, $s_1=\sqrt{2(\lambda+\mu)}$,
$s_2=\sqrt{2\mu}$, and shear entry $s_2$.  Each normal row of
$\tilde{\mathbf{C}}^{1/2}/\sqrt\rho$ has Euclidean norm exactly $c_p$, and the shear
row enters $\mathbf{G}$ with weight $\sqrt{2\mu}/\sqrt{2\rho}=c_s$.  The root is
real whenever $\tilde{\mathbf{C}}$ is positive semidefinite
($\lambda\ge-2\mu/k$); our implementation assumes $\lambda\ge0$
($c_p\ge\sqrt2\,c_s$).

On the grid each $\partial$ in $\mathbf{E}$ becomes a
forward or a backward difference, and the adjoint rows of
Eq.~\eqref{eq:Helastic} then contain the opposite member automatically
through Eq.~\eqref{eq:adjoint}.  Every assignment gives a Hermitian $\mathbf{H}$.
We take the classical staggered one: backward differences for the normal
(diagonal) strain rates and forward differences for the shear,
\begin{equation}
  \mathbf{E}\;\longrightarrow\;
  \begin{bmatrix}
    \Db_x & 0\\ 0 & \Db_y\\ \Df_y/\sqrt2 & \Df_x/\sqrt2
  \end{bmatrix}.
  \label{eq:stagpair}
\end{equation}
It corresponds to the staggered positions~\cite{virieux1986psv}: $\tilde v_x$ at $(i{+}\tfrac12,j)$,
$\tilde v_y$ at $(i,j{+}\tfrac12)$, $s_{xx},s_{yy}$ at the nodes, and
$s_{xy}$ at $(i{+}\tfrac12,j{+}\tfrac12)$.  With $\mathbf{E}$ replaced
by Eq.~\eqref{eq:stagpair}, $\mathbf{G}$ becomes a $3\times2$ array of
blocks $\mathbf{G}_{ab}$ (stress $a$, velocity $b$), and the discrete
elastic Hamiltonian, the counterpart of Eq.~\eqref{eq:Hdisc}, is
\begin{equation}
  \mathbf{H} = i\sum_{a=1}^{3}\sum_{b=1}^{2}
  \Bigl(\ket{a+1}\!\bra{b-1}\otimes\mathbf{G}_{ab}
      - \ket{b-1}\!\bra{a+1}\otimes\mathbf{G}_{ab}^{\dagger}\Bigr),
  \label{eq:Helasticdisc}
\end{equation}
where $\ket{b'}\!\bra{b''}$ acts on the block label ($0,1$ for
$\tilde v_x,\tilde v_y$ and $2,3,4$ for $s_{xx},s_{yy},s_{xy}$).

The truncated differences close the domain
exactly as in the acoustic case (ghost values zero beyond the register),
so every edge again reflects totally and no energy leaves.

In one dimension, for waves that depend on $x$ only, the
velocity--stress system splits into a P pair $(v_x,\sigma_{xx})$ with
speed $c_p$ and an S pair $(v_y,\sigma_{xy})$ with speed $c_s$ (and a
second S pair $(v_z,\sigma_{xz})$ in three-component motion).  Each pair
is a copy of the one-dimensional acoustic system: with $\tilde v=\sqrt\rho\,v$
and $s=\sigma/(\sqrt\rho\,c)$, $\partial_t\tilde v=c\,\partial_xs$ and
$\partial_ts=c\,\partial_x\tilde v$, i.e.\ Eq.~\eqref{eq:firstorder} with
$q=is$.  Discretely the pairing of Eq.~\eqref{eq:stagpair} is restricted to
$\partial_x$.
In three dimensions the Voigt vector is
$\tilde{\boldsymbol{\upsigma}}=(\sigma_{xx},\sigma_{yy},\sigma_{zz},\sqrt2\sigma_{yz},
\sqrt2\sigma_{xz},\sqrt2\sigma_{xy})$, $\tilde{\mathbf{C}}^{1/2}$ is
Eq.~\eqref{eq:csqrt} with $k=3$, and $\mathbf{E}$ has the rows
$(\partial_x,0,0)$, $(0,\partial_y,0)$, $(0,0,\partial_z)$,
$(0,\partial_z,\partial_y)/\sqrt2$, $(\partial_z,0,\partial_x)/\sqrt2$,
$(\partial_y,\partial_x,0)/\sqrt2$, discretized with backward
differences in the three normal rows and forward differences in the
three shear rows, the staggered grid of Graves~\cite{graves1996}.
Nothing else in the derivation depends on the dimension.

In one dimension the P--SV system fills four blocks without padding,
$[\tilde{\mathbf{v}}_x;\tilde{\mathbf{v}}_y;\mathbf{s}_{xx};\mathbf{s}_{xy}]$ on $n_x+2$ qubits.
In two dimensions the five fields
$[\tilde{\mathbf{v}}_x;\tilde{\mathbf{v}}_y;\mathbf{s}_{xx};\mathbf{s}_{yy};\mathbf{s}_{xy}]$ are padded to eight
blocks: $n_x+n_y+3$ qubits.  In three dimensions the nine fields are
padded to sixteen blocks: $n_x+n_y+n_z+4$ qubits.

\subsection{Summary of the encodings}

Table~\ref{tab:qubits} collects the register sizes.  In every case the
undamped semi-discrete dynamics is $\dot{\boldsymbol{\uppsi}}=-i\mathbf{H}\boldsymbol{\uppsi}$ with $\mathbf{H}$ Hermitian
and block off-diagonal between the ``velocity'' and the
``gradient/stress'' fields.  The acoustic $\mathbf{H}$ of Eq.~\eqref{eq:Hdisc} is
real symmetric (the factor $i$ sits in the field $\mathbf{q}$); the elastic $\mathbf{H}$ of
Eq.~\eqref{eq:Helastic} is $i\mathbf{A}_0$ with $\mathbf{A}_0$ real antisymmetric.

\begin{table}[tbp]
  \caption{Register sizes of the encodings ($N_a=2^{n_a}$ grid points per
  axis).  The Schr\"odingerized simulation adds $n_p$ qubits for the
  auxiliary variable ($n_p=9$--$11$ in our gate-level runs) and, optionally,
  one flag ancilla for the structured damping circuit.}
  \label{tab:qubits}
  \centering
  \begin{tabular}{lccc}
    \toprule
    medium & fields & blocks & system qubits \\
    \midrule
    acoustic 1D & $v,q$ & 2 & $n_x+1$\\
    acoustic 2D & $v,q_x,q_y$ & 4 & $n_x+n_y+2$\\
    acoustic 3D & $v,q_x,q_y,q_z$ & 4 & $n_x+n_y+n_z+2$\\
    elastic 1D  & $\tilde v_{x,y}$, $s_{xx,xy}$ & 4 & $n_x+2$\\
    elastic 2D  & $\tilde v_{x,y}$, $s_{xx,yy,xy}$ & 8 & $n_x+n_y+3$\\
    elastic 3D  & $\tilde v_{x,y,z}$, 6 stresses & 16 & $n_x+n_y+n_z+4$\\
  \bottomrule
  \end{tabular}
\end{table}

%% file: sections/sponge.tex
\section{The sponge layer: a dissipative generator}
\label{sec:sponge}

\subsection{Continuum}

Surround the region of interest by layers of thickness $L$ on every
side and damp all wave fields there at a local rate
$\sigma(x)\ge0$:
\begin{equation}
  \partial_t\boldsymbol{\uppsi} = -i\mathbf{H}\boldsymbol{\uppsi} - \boldsymbol{\Sigma}(x)\,\boldsymbol{\uppsi},
  \qquad
  \sigma(x)=\sum_{a=1}^{d}\sigma_a(x_a),
  \label{eq:spongecont}
\end{equation}
where $\boldsymbol{\Sigma}(x)$ multiplies each component of $\boldsymbol{\uppsi}$ by $\sigma(x)$ and
$\sigma_a$ vanishes outside the two layers normal to axis $a$.  In a
corner region the rates of the two (or three) layers add.

Written out in physical variables, the damped acoustic system for the
real fields $v=\partial_tu$ and $\mathbf{r}=c\nabla u$ ($=-i\mathbf{q}$) reads
\begin{equation}
  \partial_t v = c\,\nabla\!\cdot\mathbf{r} - \sigma(x)\,v,
  \qquad
  \partial_t\mathbf{r} = c\,\nabla v - \sigma(x)\,\mathbf{r} ,
  \label{eq:dampedacoustic}
\end{equation}
i.e.\ the undamped first-order system of Eq.~\eqref{eq:firstorder} with a
friction term on each field.  Where $\sigma=0$ it is again the wave
equation, Eq.~\eqref{eq:wave}; inside the layer $\mathbf{r}$ is no longer the
gradient of a single potential, so the first-order system, not
Eq.~\eqref{eq:wave}, is the model.  For the elastic system, in the
scaled variables of Eq.~\eqref{eq:elasticpair},
\begin{equation}
  \partial_t\mathbf{s} = \mathbf{G}\tilde{\mathbf{v}} - \sigma(x)\,\mathbf{s},
  \qquad
  \partial_t\tilde{\mathbf{v}} = -\mathbf{G}^{\dagger}\mathbf{s} - \sigma(x)\,\tilde{\mathbf{v}} ,
  \label{eq:dampedelastic}
\end{equation}
so particle velocity and stresses are damped at the same local rate.

\begin{remark}
\label{rem:mask}
Second-order codes usually add the sponge as a single friction term,
$u_{tt}+\gamma(x)\,u_t=c^{2}\Delta u$, which in first-order form
damps $v$ alone: $\partial_tv=c\nabla\!\cdot\mathbf{r}-\gamma v$,
$\partial_t\mathbf{r}=c\nabla v$.  Eq.~\eqref{eq:dampedacoustic} damps both
fields.  For constant $\sigma$, applying $(\partial_t+\sigma)$ twice
gives
\begin{equation}
  v_{tt}+2\sigma v_t+\sigma^{2}v=c^{2}\Delta v ,
  \label{eq:secondorder}
\end{equation}
whose solution is $e^{-\sigma t}$ times the undamped wave: the pulse is
attenuated without changing shape.  The mask with $\gamma=2\sigma$ has
the same friction term but lacks $\sigma^{2}v$.  It is then a telegraph
equation whose solutions spread, and with a graded $\gamma(x)$ its
characteristic families couple even in one dimension, so the mask
reflects where Eq.~\eqref{eq:dampedacoustic} does not.  It also leaves
the gradient field undamped, so a slowly decaying remnant stays behind.
Both versions have a diagonal, negative semidefinite Hermitian part and
fit the quantum algorithm equally well; we compare
them numerically below, for the acoustic and the elastic system (where the
mask damps the velocities only).
\end{remark}
Equation~\eqref{eq:spongecont} is the
sponge of Israeli and Orszag~\cite{israeli1981} applied to the first-order
system, and a
complex absorbing potential for the Schr\"odinger form
$i\partial_t\boldsymbol{\uppsi}=(\mathbf{H}-i\boldsymbol{\Sigma})\boldsymbol{\uppsi}$.  The energy is no longer conserved:
\begin{equation}
  \frac{d}{dt}\|\boldsymbol{\uppsi}\|^{2}
  = 2\,\mathrm{Re}\,\langle\boldsymbol{\uppsi},(-i\mathbf{H}-\boldsymbol{\Sigma})\boldsymbol{\uppsi}\rangle
  = -2\,\langle\boldsymbol{\uppsi},\boldsymbol{\Sigma}\boldsymbol{\uppsi}\rangle \;\le\;0 ,
  \label{eq:energydecay}
\end{equation}
because $\langle\boldsymbol{\uppsi},-i\mathbf{H}\boldsymbol{\uppsi}\rangle$ is purely imaginary.  Energy leaves
only where $\sigma>0$, at the rate $2\sigma$ times the local energy
density.

In $d=1$ write
the real fields $v=\partial_tu$ and $r=c\,\partial_xu=-iq$.
Equation~\eqref{eq:spongecont} reads $\partial_tv=c\,\partial_xr-\sigma
v$, $\partial_tr=c\,\partial_xv-\sigma r$, and the characteristic
variables $a_\pm=v\pm r$ obey
\begin{equation}
  (\partial_t\mp c\,\partial_x+\sigma(x))\,a_\pm=0 .
  \label{eq:characteristics}
\end{equation}
The two families stay decoupled for any $\sigma(x)$: a wave
entering the layer is attenuated along its characteristic and never
generates a reflected wave.  The only reflection is the one from the
outer edge, after which the wave crosses the
layer a second time.  Along a characteristic $dx=c\,dt$, so a wave that
traverses the layer and returns is attenuated by
$\exp(-2\int_0^{L}\sigma\,dx/c)$.  For the polynomial profile
\begin{equation}
  \sigma_a(x_a)=\sigma_{\max}\Bigl(\frac{\delta(x_a)}{L}\Bigr)^{m},
  \label{eq:profile}
\end{equation}
with $\delta$ the depth into the layer, this round-trip attenuation is
$R_0=\exp[-2\sigma_{\max}L/((m+1)c)]$, i.e.\ the amplitude
\begin{equation}
  \sigma_0 = \frac{(m+1)\,c\,\ln(1/R_0)}{2L}
  \label{eq:sigma0}
\end{equation}
delivers a prescribed round-trip attenuation $R_0$, the standard design
rule for graded absorbing layers~\cite{collino2001pml}.  We use $m=2$ and
$R_0=10^{-3}$, take $\sigma_0$ as the reference amplitude and vary
$\sigma_{\max}/\sigma_0$.

For $d\ge2$ the decoupling fails at oblique
incidence: an obliquely incident wave carries a tangential gradient that
is damped together with the normal one, and the layer reflects.  The
reflection is small when the profile varies slowly on the scale of a
wavelength and grows with the steepness of the profile.  A weak layer,
in turn, lets the wave cross it, reflect off the outer edge and come
back.  This trade-off is the well-known limitation of
sponges~\cite{israeli1981,bodony2006analysis}.  The sponge is attractive here for a
different reason: its dissipation is local and diagonal, and this makes
the quantum algorithm clean.

\subsection{Discretization}

On the grid each field is damped with the profile sampled at its own
position.  In one dimension
$\boldsymbol{\Sigma}=\diag(\sigma(x_j),\,\sigma(x_{j-1/2}))$ for $[\mathbf{v};\mathbf{q}]$.  For the acoustic
system in two dimensions,
\begin{equation}
  \boldsymbol{\Sigma} = \diag\bigl(\boldsymbol{\Sigma}^{(v)},\,\boldsymbol{\Sigma}^{(q_x)},\,\boldsymbol{\Sigma}^{(q_y)},\,\mathbf{0}\bigr),
  \label{eq:Sigmadisc}
\end{equation}
where each block is diagonal, with entries $\sigma^{(v)}_{ij}=\sigma_x(x_i)+\sigma_y(y_j)$,
$\sigma^{(q_x)}_{ij}=\sigma_x(x_{i-1/2})+\sigma_y(y_j)$ and
$\sigma^{(q_y)}_{ij}=\sigma_x(x_i)+\sigma_y(y_{j-1/2})$; the padding block
is not damped.  In three dimensions $\sigma=\sigma_x+\sigma_y+\sigma_z$
with the analogous staggered sampling.  For the elastic system we damp
the five physical blocks with the nodal sum $\sigma_x(x_i)+\sigma_y(y_j)$
and leave the three padding blocks undamped.  The semi-discrete sponge
system is
\begin{equation}
  \dot{\boldsymbol{\uppsi}} = \mathbf{A}\boldsymbol{\uppsi},\qquad \mathbf{A} = -i\mathbf{H}-\boldsymbol{\Sigma} .
  \label{eq:A}
\end{equation}

Every non-unitary linear system can be split into a Hermitian and an
anti-Hermitian part, $\mathbf{A}=\mathbf{H}_1+i\mathbf{H}_2$ with
$\mathbf{H}_1=(\mathbf{A}+\mathbf{A}^{\dagger})/2$ and $\mathbf{H}_2=(\mathbf{A}-\mathbf{A}^{\dagger})/(2i)$, both Hermitian.
For the sponge the split is immediate:
\begin{equation}
  \mathbf{H}_1 = -\boldsymbol{\Sigma},\qquad \mathbf{H}_2 = -\mathbf{H} .
  \label{eq:split}
\end{equation}

\begin{lemma}[Dissipativity]
\label{lem:dissipative}
For the acoustic and elastic sponge generators of Eq.~\eqref{eq:A}, in one, two
or three dimensions and for any nonnegative profiles, $\mathbf{H}_1=-\boldsymbol{\Sigma}$ is
diagonal and negative semidefinite.  Consequently
$\|e^{\mathbf{A}t}\|\le1$ for all $t\ge0$,
$\frac{d}{dt}\|\boldsymbol{\uppsi}\|^{2}=-2\langle\boldsymbol{\uppsi},\boldsymbol{\Sigma}\boldsymbol{\uppsi}\rangle$, and
$\lamax^{+}(\mathbf{H}_1):=\max(0,\lamax(\mathbf{H}_1))=0$.
\end{lemma}

\begin{proof}
$\boldsymbol{\Sigma}$ is diagonal with entries $\sigma\ge0$, so $-\boldsymbol{\Sigma}\preceq0$.
The energy identity is Eq.~\eqref{eq:energydecay} with the discrete $\mathbf{H}$,
which is Hermitian by Eq.~\eqref{eq:adjoint}.  It gives
$\|e^{\mathbf{A}t}\boldsymbol{\uppsi}_0\|\le\|\boldsymbol{\uppsi}_0\|$.
\end{proof}

Lemma~\ref{lem:dissipative} is the structural fact behind Schr\"odingerization
of the sponge: $\lamax^{+}(\mathbf{H}_1)$ is the quantity that controls
the Schr\"odingerization recovery~\cite{jin2024recovery}, and for the
sponge it vanishes.  The largest damping rate
\begin{equation}
  \sbar = \max\boldsymbol{\Sigma} = \lVert \mathbf{H}_1\rVert
  \label{eq:sbar}
\end{equation}
($\sbar=d\,\sigma_{\max}$, attained in the corners) will set the speed
of the auxiliary dynamics.

\subsection{Classical validation}
\label{sec:validation}

We damp both $v$ and $\mathbf{r}$ [Eqs.~\eqref{eq:dampedacoustic} and
\eqref{eq:dampedelastic}].  Before turning to the quantum algorithm we
verify, with purely classical computations, that this is a good
absorber: first against the alternatives (damping $v$ only, no damping)
with an independent finite-difference code, then as a function of the
layer parameters.  Unless stated otherwise, $h=1$, $c=1$
(acoustic), and $c_p=1$, $c_s=1/\sqrt3$, $\rho=1$ (elastic).

\subsubsection{Damping \texorpdfstring{$v$ and $\mathbf{r}$}{v and r} versus $v$ only}
\label{sec:fdcheck}

\begin{figure}[tbp]
  \centering
  \includegraphics[width=\textwidth]{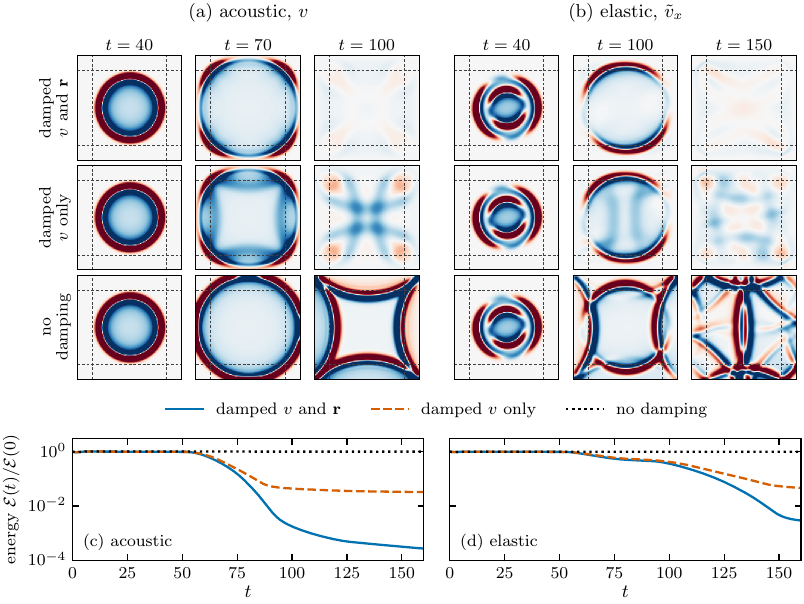}
  \caption{Independent classical (leapfrog finite-difference) solutions
  of the damped wave equations.  Rows: damped $v$ and $\mathbf{r}$
  [Eqs.~\eqref{eq:dampedacoustic} and \eqref{eq:dampedelastic}; for the
  elastic system ``$\mathbf{r}$'' stands for the stresses]; damped $v$ only,
  the mask $u_{tt}+\gamma u_t=c^{2}\Delta u$ with $\gamma=2\sigma$; no
  damping.  Left: acoustic $v$ at $t=40,70,100$.  Right: elastic
  $\tilde v_x$ (Virieux scheme, $c_p=1$, $c_s=1/\sqrt3$, $\rho=1$, so $\tilde v_x=v_x$) at
  $t=40,100,150$, later because the slower S front needs longer to
  cross the layer.
  Colour clipped at $25\%$ of the peak of each system; dashed lines:
  layer edges.  Bottom: energy remaining in the domain; the elastic curve
  drops in two stages as the P and then the S wave leave.}
  \label{fig:fd}
\end{figure}

We solve Eq.~\eqref{eq:dampedacoustic} with a standard staggered-grid
leapfrog finite-difference scheme, written independently of the
qubit-encoding code
(time step $\Delta t=0.4h/c$, semi-implicit damping, $100\times100$
interior, $20$-node layers, $\sigma_{\max}=\sigma_0$, Gaussian pulse of
width $3h$).  We compare damping both $v$ and $\mathbf{r}$
[Eq.~\eqref{eq:dampedacoustic}] with damping $v$ only
(Remark~\ref{rem:mask}, $\gamma=2\sigma$) and with no damping, each
against an absorber-free reference on a three times larger grid
(Fig.~\ref{fig:fd} and Table~\ref{tab:fd}).  Damping $v$ and $\mathbf{r}$
absorbs the outgoing ring between $t=70$ and $t=100$ and leaves a faint
residual;
damping $v$ only reflects $5.5$ times more and keeps $3\%$ of the energy in the
domain; without damping nothing leaves.  The elastic counterpart, solved with an
independent Virieux staggered-grid leapfrog scheme ($c_p=1$,
$c_s=1/\sqrt3$, same layer and pulse, initial velocity
$(\tilde v_x,\tilde v_y)\propto(1,0.5)$), behaves the same way
[Eq.~\eqref{eq:dampedelastic} damps velocities and stresses]: damping
velocities only reflects $3.5$ times more and leaves $16$ times more
energy, and the energy curve shows the P and then the slower S wave
leaving the domain (Fig.~\ref{fig:fd}, bottom right).  An amplitude scan gives the
same optimum, $\sigma_{\max}\approx\sigma_0/2$, and the same
$3\times10^{-2}$ reflection level as Fig.~\ref{fig:absorption}, which is computed with the qubit-encoded
operator and matrix exponentials: the two codes agree on the physics of
the layer.

\begin{table}[tbp]
  \caption{Classical finite-difference comparison of absorbers
  (Fig.~\ref{fig:fd}): maximum reflected amplitude in the interior
  window over $t\le160$ (relative to the initial pulse) and energy left
  in the domain at $t=160$.  Elastic: $\mathbf{r}$ stands for the stresses
  [Eq.~\eqref{eq:dampedelastic}].}
  \label{tab:fd}
  \centering
  \begin{tabular}{lcc}
    \toprule
    damping & max reflection & energy left\\
    \midrule
    \multicolumn{3}{l}{acoustic}\\
    damped $v$ and $\mathbf{r}$, Eq.~\eqref{eq:dampedacoustic} & $2.8\times10^{-2}$ & $2.7\times10^{-4}$\\
    damped $v$ only & $1.5\times10^{-1}$ & $3.2\times10^{-2}$\\
    no damping (rigid walls) & $6.0\times10^{-1}$ & $1.0$\\
    \multicolumn{3}{l}{elastic}\\
    damped $v$ and $\mathbf{r}$, Eq.~\eqref{eq:dampedelastic} & $4.7\times10^{-2}$ & $2.9\times10^{-3}$\\
    damped $v$ only & $1.7\times10^{-1}$ & $4.6\times10^{-2}$\\
    no damping (rigid walls) & $4.6\times10^{-1}$ & $1.0$\\
  \bottomrule
  \end{tabular}
\end{table}

\subsubsection{Layer width and amplitude}
\label{sec:absorption}

\begin{figure}[tbp]
  \centering
  \includegraphics[width=\textwidth]{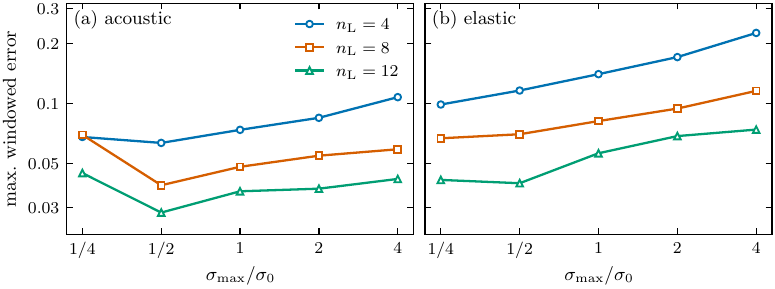}
  \caption{Spurious reflection of the sponge layer on a $32\times32$
  grid: maximum windowed error against a fourfold enlarged
  absorber-free reference, versus the damping amplitude relative to the
  one-dimensional design value $\sigma_0$ of Eq.~\eqref{eq:sigma0}, for
  layer widths $\nL=4,8,12$.}
  \label{fig:absorption}
\end{figure}

We next measure the reflection as a function of layer width and damping
amplitude, on the discrete system of Eq.~\eqref{eq:A} itself (the operator
that the quantum algorithm simulates).  A $32\times32$ sponge-terminated
run is compared
with the same initial pulse (Gaussian, width $1.5$, $\|\boldsymbol{\uppsi}_0\|=1$)
evolved on a $128\times128$ grid without any absorber.  The comparison
is made on the interior window (layers excluded) up to $T=40$, before
the large grid's own edge reflections can return.  The windowed
difference is the spurious reflection from the sponge.
Figure~\ref{fig:absorption} shows its maximum over
$T\in\{10,20,30,40\}$ as a function of the amplitude
$\sigma_{\max}/\sigma_0$ [Eq.~\eqref{eq:sigma0}] for layer widths
$\nL=4,8,12$.

The acoustic sponge shows the classic trade-off: a weak layer
($\sigma_{\max}=\sigma_0/4$) lets the pulse cross it, reflect off the
outer edge and return (energy left at $T=40$: $3$--$4\times10^{-3}$);
a strong layer reflects off its own gradient.  The optimum lies at
$\sigma_{\max}\approx\sigma_0/2$, with reflected amplitudes
$6.3\times10^{-2}$, $3.9\times10^{-2}$ and $2.8\times10^{-2}$ for
$\nL=4,8,12$.  Widening the layer helps steadily, as expected for a
gradient-limited absorber; at the optimum $9\times10^{-5}$--$8\times10^{-4}$
of the energy remains in the domain at $T=40$.  (The grid is fixed, so
the comparison window shrinks from $24\times24$ to $8\times8$ nodes as
$\nL$ grows from 4 to 12; the three widths are not measured on the same
window.)  The elastic sponge
reflects more ($9.9\times10^{-2}$, $6.7\times10^{-2}$ and
$4.0\times10^{-2}$ at its best amplitude for $\nL=4,8,12$) and prefers
weaker damping (for $\nL=4,8$ the best value is the weakest amplitude
scanned, so the optimum is not bracketed), presumably because the slower S waves see a steeper
layer in units of their own wavelength.  These percent-level reflections are the price
of the simplest dissipative absorber.  What the sponge buys is the
clean quantum structure.
Wider layers are cheap on a quantum register: doubling $\nL$ at fixed
interior costs at most one qubit per axis.

%% file: sections/schro.tex
\section{Schr\"odingerization of the sponge}
\label{sec:schro}

We now turn the dissipative system $\dot{\boldsymbol{\uppsi}}=\mathbf{A}\boldsymbol{\uppsi}$,
$\mathbf{A}=\mathbf{H}_1+i\mathbf{H}_2=-\boldsymbol{\Sigma}-i\mathbf{H}$, into a unitary evolution.  The construction
is that of Jin \emph{et al.}~\cite{jin2024prl,jin2023schrodingerisation}; we derive
it step by step and specialize every statement to the sponge, for which
it becomes particularly transparent.

\subsection{The warped phase transformation}
\label{sec:warped}

Introduce an auxiliary variable $p\in\mathbb R$ and, for $p>0$,
\begin{equation}
  \mathbf{w}(t,p) = e^{-p}\,\boldsymbol{\uppsi}(t).
  \label{eq:warp}
\end{equation}
Differentiating, $\partial_t \mathbf{w}=e^{-p}\mathbf{A}\boldsymbol{\uppsi}=\mathbf{H}_1e^{-p}\boldsymbol{\uppsi}+i\mathbf{H}_2\mathbf{w}$, and
$e^{-p}\boldsymbol{\uppsi}=-\partial_p\mathbf{w}$.  Hence
\begin{equation}
  \partial_t \mathbf{w} = -\mathbf{H}_1\,\partial_p \mathbf{w} + i\mathbf{H}_2\,\mathbf{w}
  \;\overset{\text{sponge}}{=}\;
  \boldsymbol{\Sigma}\,\partial_p \mathbf{w} - i\mathbf{H}\,\mathbf{w} .
  \label{eq:warpedpde}
\end{equation}
The damping, a zeroth-order term in $\boldsymbol{\uppsi}$, has become a first-order
transport term in $p$, with the diagonal matrix $\boldsymbol{\Sigma}$ as the
velocity.  The anti-Hermitian part $-i\mathbf{H}$ is untouched.
Equation~\eqref{eq:warpedpde} is posed on all of $\mathbb R$ with the
initial data
\begin{equation}
  \mathbf{w}(0,p) = g(p)\,\boldsymbol{\uppsi}_0,
  \qquad g(p)=e^{-p}\ \text{for } p\ge0,
  \label{eq:warpinit}
\end{equation}
where $g$ on $p<0$ is a free choice to be made for numerical
convenience.  The solution is a function of
$(t,p)$ with values in the state space of $\boldsymbol{\uppsi}$; the original solution
is read off on $p\ge0$.

\begin{example}[One damped mode]
\label{ex:scalar}
Let $\boldsymbol{\uppsi}$ have a single component (one complex amplitude), with $\mathbf{A}=-\sigma-i\omega$
($\sigma\ge0$).  Equation~\eqref{eq:warpedpde} is
$\partial_t\mathbf{w}=\sigma\partial_p\mathbf{w}-i\omega \mathbf{w}$, whose solution is
$\mathbf{w}(t,p)=e^{-i\omega t}g(p+\sigma t)\boldsymbol{\uppsi}_0$: the initial profile is
transported rigidly to the left at speed $\sigma$ while it
oscillates.  For $p\ge0$ the argument $p+\sigma t$ is nonnegative, so
$\mathbf{w}(t,p)=e^{-i\omega t}e^{-p-\sigma t}\boldsymbol{\uppsi}_0=e^{-p}\boldsymbol{\uppsi}(t)$, exactly the
warped solution.  Damping is encoded as a shift: the profile that sits
at $p$ at time $t$ started at $p+\sigma t$, where its height was smaller
by $e^{-\sigma t}$.  The value of $g$ on $p<0$ never reaches $p\ge0$.
\end{example}

For the full system $\boldsymbol{\Sigma}$ and $\mathbf{H}$ do not commute, so there is no
rigid shift, but the two conclusions of Example~\ref{ex:scalar}
survive.

\begin{proposition}[Exact recovery on the half-line]
\label{prop:exact}
Let $\mathbf{H}_1=-\boldsymbol{\Sigma}\preceq0$ and let $\mathbf{w}$ solve
Eqs.~\eqref{eq:warpedpde}--\eqref{eq:warpinit} with any $g\in L^{2}$
that equals $e^{-p}$ on $p\ge0$.  Then for every $t\ge0$,
\begin{equation}
  \mathbf{w}(t,p)=e^{-p}\boldsymbol{\uppsi}(t)\qquad\text{for all } p\ge0 .
  \label{eq:exact}
\end{equation}
In particular $\boldsymbol{\uppsi}(t)=e^{p^{*}}\mathbf{w}(t,p^{*})$ for every $p^{*}\ge0$,
$\boldsymbol{\uppsi}(t)=\int_0^{\infty}\mathbf{w}(t,p)\,dp$, and
\begin{equation}
  \int_0^{\infty}\|\mathbf{w}(t,p)\|^{2}\,dp=\tfrac12\|\boldsymbol{\uppsi}(t)\|^{2}.
  \label{eq:normidentity}
\end{equation}
\end{proposition}

\begin{proof}
$e^{-p}\boldsymbol{\uppsi}(t)$ solves Eq.~\eqref{eq:warpedpde} on all of $\mathbb R$
(the computation above did not use $p>0$).  The difference
$\mathbf{e}(t,p)=\mathbf{w}-e^{-p}\boldsymbol{\uppsi}(t)$ therefore solves Eq.~\eqref{eq:warpedpde} with
initial data $(g(p)-e^{-p})\boldsymbol{\uppsi}_0$, which vanishes on $p\ge0$.  For
$a\ge0$,
\begin{equation*}
\begin{split}
  \frac{d}{dt}\int_a^{\infty}\|\mathbf{e}\|^{2}dp
  &= 2\,\mathrm{Re}\!\int_a^{\infty}\!\langle\mathbf{e},\boldsymbol{\Sigma}\partial_p\mathbf{e}\rangle dp\\
  &= -\langle\mathbf{e},\boldsymbol{\Sigma}\mathbf{e}\rangle\big|_{p=a}\le0,
\end{split}
\end{equation*}
because $\boldsymbol{\Sigma}$ is Hermitian, $\langle\mathbf{e},-i\mathbf{H}\mathbf{e}\rangle$ is imaginary,
and $\boldsymbol{\Sigma}\succeq0$.  The integral vanishes at $t=0$ and is nonnegative,
so it vanishes for all $t$.  The remaining statements follow from
$\int_0^{\infty}e^{-p}dp=1$ and $\int_0^{\infty}e^{-2p}dp=\tfrac12$.
\end{proof}

For a general generator the same argument requires the boundary term
$+\langle\mathbf{e},\mathbf{H}_1\mathbf{e}\rangle|_{p=a}$ to be nonpositive, which fails when
$\mathbf{H}_1$ has positive eigenvalues; recovery is then valid only for
$p^{*}\ge\lamax^{+}(\mathbf{H}_1)\,t$~\cite{jin2024recovery}.  The sponge has
$\lamax^{+}(\mathbf{H}_1)=0$ (Lemma~\ref{lem:dissipative}), so any $p^{*}\ge0$
works at every time.

\subsection{Characteristics and the domain of dependence}
\label{sec:dod}

Equation~\eqref{eq:warpedpde} is a symmetric hyperbolic system in $p$
with characteristic speeds equal to minus the eigenvalues of $\boldsymbol{\Sigma}$:
all information moves to the left (or stands still), at most at speed
$\sbar=\max\boldsymbol{\Sigma}$.  The value of $\mathbf{w}$ at $(t,p)$ is therefore determined
by the initial data on $[p,\,p+\sbar t]$.  This has a practical
consequence for the truncation of the $p$ axis.

\begin{proposition}[Domain of dependence]
\label{prop:dod}
Let Eq.~\eqref{eq:warpedpde} be posed on the periodic interval
$[-p_{\max},p_{\max})$ with initial data $g(p)\boldsymbol{\uppsi}_0$, $g=e^{-p}$ on
$[0,p_{\max})$.  If $\sbar t\le p_{\max}$, then
$\mathbf{w}(t,p)=e^{-p}\boldsymbol{\uppsi}(t)$ for all $p\in[0,\,p_{\max}-\sbar t]$.  In
particular, recovery at the slice $p^{*}$ is exact as long as
\begin{equation}
  p_{\max}\;\ge\;p^{*}+\sbar\,T .
  \label{eq:pmaxrule}
\end{equation}
\end{proposition}

\begin{proof}
Let $\mathbf{e}=\mathbf{w}-e^{-p}\boldsymbol{\uppsi}(t)$ on the arc $(0,p_{\max})$, where both terms are
smooth solutions; $\mathbf{e}(0,\cdot)=0$ there.  With $b(t)=p_{\max}-\sbar t$
and $0\le a<b(t)$,
\begin{equation*}
  \frac{d}{dt}\int_a^{b(t)}\!\|\mathbf{e}\|^{2}dp
  = -\langle\mathbf{e},\boldsymbol{\Sigma}\mathbf{e}\rangle\big|_{a}
    -\bigl\langle\mathbf{e},(\sbar-\boldsymbol{\Sigma})\mathbf{e}\bigr\rangle\big|_{b(t)}\le0,
\end{equation*}
since $\boldsymbol{\Sigma}\succeq0$ and $\sbar\mathbf{I}-\boldsymbol{\Sigma}\succeq0$; the second term
collects the flux $+\langle\mathbf{e},\boldsymbol{\Sigma}\mathbf{e}\rangle$ through the right end and
the shrinking of the interval, $b'(t)\|\mathbf{e}\|^{2}=-\sbar\|\mathbf{e}\|^{2}$.  Hence
$\mathbf{e}\equiv0$ on $[a,b(t)]$.
\end{proof}

Beyond the time $T_{\rm dep}=(p_{\max}-p^{*})/\sbar$ the slice
$p^{*}$ starts to receive data that entered through the periodic seam at
$p_{\max}$, i.e.\ the values $g(p-2p_{\max})$ on the negative side
instead of $e^{-p}$.  A component that has been damped at rate $\sbar$
for the whole time is then contaminated at the level
$e^{p^{*}}g(p^{*}+\sbar T-2p_{\max})\approx e^{2p^{*}+\sbar T-2p_{\max}}$
relative to $\|\boldsymbol{\uppsi}_0\|$ (for $g=e^{-|p|}$ on the negative side).  The
bound of Eq.~\eqref{eq:pmaxrule} uses the worst rate $\sbar$, attained only in
the corners of the layer.  A wave spends only a short time there before
it is absorbed, and in practice the contamination sets in later.  Because $p_{\max}$ enters the qubit count only
through $\log_2(p_{\max}/\Delta p)$, Eq.~\eqref{eq:pmaxrule} costs
$\log_2(\sbar T)$ qubits.

\begin{remark}
\label{rem:alpha}
Choosing $g(p)=e^{-\alpha|p|}$ with $\alpha>1$ on $p<0$, as in
\citet{jin2023schrodingerisation}, here joined to $e^{-p}$ by a cubic
Hermite piece, makes the seam data smaller and
delays the contamination; a $C^{1}$ junction is still required for
accuracy.  On the $8\times8$ acoustic sponge with $p_{\max}=8$, $\alpha=4$ lowers the error at
$T=8$ from $9.2\times10^{-4}$ to $6.2\times10^{-5}$ (absolute, relative
to $\|\boldsymbol{\uppsi}_0\|$) but does not move the
time at which the corner content has wrapped once around the periodic
interval.
\end{remark}

\subsection{Fourier transform in $p$: a family of Hamiltonians}
\label{sec:fourier}

Transport in $p$ is diagonalized by the Fourier transform.  With
$\hat{\mathbf{w}}(t,\eta)=\int \mathbf{w}(t,p)e^{-i\eta p}\,dp$ the derivative $\partial_p$
becomes multiplication by $i\eta$, and Eq.~\eqref{eq:warpedpde} becomes
\begin{equation}
  i\,\partial_t\hat{\mathbf{w}} = (\eta \mathbf{H}_1 - \mathbf{H}_2)\,\hat{\mathbf{w}}
  = \bigl(\mathbf{H}-\eta\,\boldsymbol{\Sigma}\bigr)\hat{\mathbf{w}} .
  \label{eq:hamfamily}
\end{equation}
For every real $\eta$ the operator $\mathbf{H}(\eta)=\mathbf{H}-\eta\boldsymbol{\Sigma}$ is Hermitian:
each Fourier mode of the warped variable is an ordinary Schr\"odinger
equation in which the damping has turned into a real potential
$-\eta\sigma(x)$, a potential barrier for $\eta<0$ and a well for
$\eta>0$.  The non-unitary sponge dynamics is the superposition of these
unitary dynamics, weighted by $\hat g(\eta)$ and recombined by the
inverse Fourier transform:
\begin{equation}
  \mathbf{w}(T,p)=\frac{1}{2\pi}\int\hat g(\eta)\,e^{i\eta p}\,
  e^{-iT(\mathbf{H}-\eta\boldsymbol{\Sigma})}\boldsymbol{\uppsi}_0\,d\eta .
  \label{eq:superposition}
\end{equation}
This is the content of Schr\"odingerization in one line, and it is
exactly what a quantum computer can do: prepare $\hat g(\eta)$ on an
auxiliary register, apply the $\eta$-controlled Hamiltonian evolution,
and Fourier transform back.

\subsection{Discretization in $p$}
\label{sec:pdisc}

Take $N_p=2^{n_p}$ points
$p_j=-p_{\max}+j\,\Delta p$, $j=0,\dots,N_p-1$, $\Delta p=2p_{\max}/N_p$,
with periodic boundary conditions.  The discrete Fourier frequencies are
\begin{equation}
  \eta_k = \frac{2\pi\,\bar k}{N_p\Delta p}=\frac{\pi}{p_{\max}}\,\bar k,
  \qquad
  \bar k\in\Bigl\{-\frac{N_p}{2},\dots,\frac{N_p}{2}-1\Bigr\},
  \label{eq:eta}
\end{equation}
where $\bar k\equiv k\pmod{N_p}$ is the signed (two's-complement)
reading of the index $k$.  Let $\mathbf{F}$ be the unitary discrete Fourier
transform, $F_{kj}=N_p^{-1/2}e^{-2\pi ijk/N_p}$.  The Fourier spectral
derivative is $\partial_p\to i\,\mathbf{P}$ with
\begin{equation}
  \mathbf{P} = \mathbf{F}^{\dagger}\mathbf{D}_\eta \mathbf{F},\qquad \mathbf{D}_\eta=\diag(\eta_0,\dots,\eta_{N_p-1}),
  \label{eq:Pmomentum}
\end{equation}
the discrete momentum operator of
\citet{jin2023schrodingerisation}.  [The factor
$e^{-2\pi ijk/N_p}=e^{-i\eta_k(p_j+p_{\max})}$ differs from
$e^{-i\eta_kp_j}$ by a $k$-dependent phase, which cancels in
$\mathbf{F}^{\dagger}\mathbf{D}_\eta \mathbf{F}$.]

Stack
$\boldsymbol{\Psi}(t)=\sum_j\ket{j}\otimes \mathbf{w}(t,p_j)$ with the $p$ index most
significant.  Equation~\eqref{eq:warpedpde} becomes
\begin{equation}
  \frac{d\boldsymbol{\Psi}}{dt}
  = -i\,\boldsymbol{\mathcal H}\,\boldsymbol{\Psi},
  \qquad
  \boldsymbol{\mathcal H} = \mathbf{I}\otimes \mathbf{H} - \mathbf{P}\otimes\boldsymbol{\Sigma} ,
  \label{eq:Hbold}
\end{equation}
a Hamiltonian system on $n_p+n_{\rm sys}$ qubits, and in the Fourier
basis, $\hat{\boldsymbol{\Psi}}=(\mathbf{F}\otimes\mathbf{I})\boldsymbol{\Psi}$,
\begin{equation}
  \hat{\boldsymbol{\mathcal H}} = (\mathbf{F}\otimes\mathbf{I})\boldsymbol{\mathcal H}(\mathbf{F}^{\dagger}\otimes\mathbf{I})
  = \mathbf{I}\otimes \mathbf{H} - \mathbf{D}_\eta\otimes\boldsymbol{\Sigma}
  = \bigoplus_{k} \mathbf{H}(\eta_k).
  \label{eq:Hhat}
\end{equation}
In the language of the general recipe,
$\hat{\boldsymbol{\mathcal H}}=\mathbf{D}_\eta\otimes \mathbf{H}_1-\mathbf{I}\otimes \mathbf{H}_2$.  The $N_p$ Fourier
modes decouple; each evolves under $\mathbf{H}(\eta_k)$ of
Eq.~\eqref{eq:hamfamily}, and
$\hat{\boldsymbol{\Psi}}(T)=e^{-iT\hat{\boldsymbol{\mathcal H}}}\hat{\boldsymbol{\Psi}}(0)$ exactly.

The spectral discretization
converges at a rate set by the smoothness of $\mathbf{w}$ in $p$, which is that of
$g$.  The kink of $e^{-|p|}$ at $p=0$ limits it to $O(\Delta p)$.  We
use the $C^{1}$ profile of \citet[Remark~4.5]{jin2024recovery},
\begin{equation}
  g(p)=\begin{cases}
    e^{-p}, & p\ge0,\\
    c_3p^{3}+c_2p^{2}-p+1, & -1<p<0,\\
    e^{p}, & p\le-1,
  \end{cases}
  \label{eq:gcubic}
\end{equation}
with $c_3=-3+3/e$ and $c_2=-5+4/e$, whose cubic piece matches value and slope at $p=0$ and $p=-1$.  It
restores $O(\Delta p^{2})$ while keeping $g=e^{-p}$ on $p\ge0$, so
Propositions~\ref{prop:exact}--\ref{prop:dod} apply unchanged.  With
$n_p=9$--$11$ and $\Delta p\approx0.03$--$0.07$ the $p$ discretization
contributes relative errors of $10^{-6}$--$10^{-4}$ in our runs.

\subsection{The quantum algorithm}
\label{sec:algorithm}

Qiskit's quantum Fourier transform acts as
\begin{equation*}
\QFT\ket{j}=N_p^{-1/2}\sum_ke^{+2\pi ijk/N_p}\ket{k},
\end{equation*}
i.e.\ $\QFT=\mathbf{F}^{\dagger}$.  The Schr\"odingerized sponge simulation is:
\begin{enumerate}
\item \emph{Prepare} the product state
  $\ket{\boldsymbol{\Psi}_0}=\ket{g}\otimes\ket{\boldsymbol{\uppsi}_0}$ with
  $\ket g\propto\sum_jg(p_j)\ket j$ on the $p$ register and
  $\ket{\boldsymbol{\uppsi}_0}=\boldsymbol{\uppsi}_0/\|\boldsymbol{\uppsi}_0\|$ on the system register.
\item Apply $\QFT^{\dagger}=\mathbf{F}$ to the $p$ register: amplitudes become
  $\hat g_k/\|g\|$, the $\eta$ basis.
\item \emph{Evolve} with $e^{-iT\hat{\boldsymbol{\mathcal H}}}$,
  $\hat{\boldsymbol{\mathcal H}}=\mathbf{I}\otimes \mathbf{H}-\mathbf{D}_\eta\otimes\boldsymbol{\Sigma}$: a Hamiltonian
  simulation controlled by the $p$ register.
\item Apply $\QFT=\mathbf{F}^{\dagger}$ to the $p$ register: back to the $p$
  basis.  The amplitude of $\ket j\otimes\ket s$ is now
  $w_s(T,p_j)/\|\boldsymbol{\Psi}_0\|$.
\item \emph{Measure} (part of) the $p$ register and keep the outcomes
  with $p_j\ge0$.
\end{enumerate}
Steps 2--4 implement
$(\mathbf{F}^{\dagger}\otimes\mathbf{I})e^{-iT\hat{\boldsymbol{\mathcal H}}}(\mathbf{F}\otimes\mathbf{I})=e^{-iT\boldsymbol{\mathcal H}}$,
the exact solution of Eq.~\eqref{eq:Hbold}.

\subsection{Measurement and success probability}
\label{sec:measure}

For the sponge every $p\ge0$ slice carries the same normalized state
(Proposition~\ref{prop:exact}), so there is no need to select a single
slice.  The half-line $p_j\ge0$ is exactly the set $j\ge N_p/2$, i.e.\
the outcome $1$ of the most significant $p$ qubit.

\begin{proposition}[One-qubit post-selection]
\label{prop:prob}
Let $J$ be a set of grid indices with $p_j\ge0$ on which
$\mathbf{w}(T,p_j)=e^{-p_j}\boldsymbol{\uppsi}(T)$.  Measuring whether $j\in J$ succeeds with
probability
\begin{equation}
  \Pr[J] = \kappa_J\,\frac{\|\boldsymbol{\uppsi}(T)\|^{2}}{\|\boldsymbol{\uppsi}_0\|^{2}},
  \qquad
  \kappa_J=\frac{\sum_{j\in J}e^{-2p_j}}{\sum_{j}g(p_j)^{2}},
  \label{eq:prob}
\end{equation}
and on success the system register is left in the pure state
$\boldsymbol{\uppsi}(T)/\|\boldsymbol{\uppsi}(T)\|$, unentangled with the $p$ register.  We write
$\kappa_g$ for $\kappa_J$ with $J=\{p_j\ge0\}$.
\end{proposition}

\begin{proof}
On success the joint state is proportional to
$\sum_{j\in J}\ket j\otimes e^{-p_j}\boldsymbol{\uppsi}(T)
=\bigl(\sum_{j\in J}e^{-p_j}\ket j\bigr)\otimes\boldsymbol{\uppsi}(T)$, a product.  Its
squared norm divided by
$\|\boldsymbol{\Psi}_0\|^{2}=\sum_jg(p_j)^{2}\|\boldsymbol{\uppsi}_0\|^{2}$ is
Eq.~\eqref{eq:prob}.
\end{proof}

For $J=\{p_j\ge0\}$ and $\Delta p\to0$, $p_{\max}\to\infty$,
$\kappa_g\to\tfrac12/\int g^{2}\,dp=0.395$ for the profile of
Eq.~\eqref{eq:gcubic} ($\tfrac12$ for $e^{-|p|}$); on our grids
$\kappa_g=0.408$--$0.424$, the excess being the $O(\Delta p)$ of the
left-endpoint sum.  The ratio $\|\boldsymbol{\uppsi}(T)\|^{2}/\|\boldsymbol{\uppsi}_0\|^{2}$
is the fraction of the initial energy that the sponge has not
absorbed.  Apart from the fixed factor $\kappa_g$, which is set by the
initial profile in $p$, the success probability is therefore determined
by the physics and not by the algorithm.  The only loss is the energy
that the sponge has removed from the domain.  Amplitude
amplification~\cite{brassard2002amplitude} reduces the number of
repetitions to
$O\bigl(\|\boldsymbol{\uppsi}_0\|/\|\boldsymbol{\uppsi}(T)\|\bigr)$.  The measured
frequency also estimates $\|\boldsymbol{\uppsi}(T)\|^{2}$, the remaining energy, which the
normalized output state cannot provide.

When the domain-of-dependence condition of Eq.~\eqref{eq:pmaxrule} is not met
with margin, the top of the $p\ge0$ window is the first to be
contaminated through the periodic seam.  Measuring
the two most significant $p$ qubits and accepting the outcome $10$
selects $p\in[0,p_{\max}/2)$, which halves the exposure at a negligible
loss of probability, since $e^{-2p}$ is concentrated near $p=0$.
Section~\ref{sec:pspace} shows that this two-qubit post-selection keeps
the output state within $4\times10^{-7}$ infidelity in a regime where the
one-qubit version reaches $10^{-1}$.

%% file: sections/circuits.tex
\section{Explicit circuits}
\label{sec:circuits}

\begin{figure}[tbp]
  \centering
  \begin{minipage}[c]{\textwidth}
  \textbf{(a)}\\[0.3ex]
  \resizebox{\linewidth}{!}{%
  \begin{quantikz}[row sep={0.95cm,between origins}, column sep=0.18cm]
    \lstick{sys} & \qwbundle{n_{\mathrm{sys}}} & \gate{\ket{\boldsymbol{\uppsi}_0}} & \qw
      & \gate[2]{e^{+i\frac{\tau}{2}\mathbf{D}_\eta\otimes\boldsymbol{\Sigma}}}
      & \gate{\mathbf{W}(\tau)}
      & \gate[2]{e^{+i\tau \mathbf{D}_\eta\otimes\boldsymbol{\Sigma}}}
      & \ \ldots\
      & \gate{\mathbf{W}(\tau)}
      & \gate[2]{e^{+i\frac{\tau}{2}\mathbf{D}_\eta\otimes\boldsymbol{\Sigma}}}
      & \qw & \qw \\
    \lstick{$p$} & \qwbundle{n_p} & \gate{\ket{g}}
      & \gate{\QFT^{\dagger}} & {} & \qw & {}
      & \ \ldots\
      & \qw & {} & \gate{\QFT} & \meter{}
  \end{quantikz}}
  \end{minipage}\\[2ex]
  \begin{minipage}[c]{0.62\textwidth}
  \textbf{(b)}\\[0.3ex]
  \resizebox{\linewidth}{!}{%
  \begin{quantikz}[row sep={0.62cm,between origins}, column sep=0.22cm]
    \lstick{$q_0\;(\sigma_{10})$} & \qw & \targ{} & \qw & \qw
      & \ctrl{1} & \qw & \qw & \targ{} & \qw & \qw \\
    \lstick{$q_1\;(\sigma_{01})$} & \targ{} & \qw & \qw & \qw
      & \octrl{3} & \qw & \qw & \qw & \targ{} & \qw \\
    \lstick{$q_2\;(I)$} & \qw & \qw & \qw & \qw
      & \qw & \qw & \qw & \qw & \qw & \qw \\
    \lstick{$q_3\;(I)$} & \qw & \qw & \qw & \qw
      & \qw & \qw & \qw & \qw & \qw & \qw \\
    \lstick{$q_4\;(\sigma_{00})$} & \qw & \qw & \qw & \qw
      & \octrl{1} & \qw & \qw & \qw & \qw & \qw \\
    \lstick{$q_5\;(\sigma_{01},\,\mathrm{piv.})$}
      & \ctrl{-4} & \ctrl{-5} & \gate{P(\phi)} & \gate{H}
      & \gate{R_z(2|\zeta|\tau)} & \gate{H} & \gate{P(-\phi)}
      & \ctrl{-5} & \ctrl{-4} & \qw
  \end{quantikz}}
  \end{minipage}\\[1ex]
  \caption{Explicit circuits.  (a)~The Schr\"odingerized sponge
  pipeline: product-state preparation, $\QFT^{\dagger}$ on the $p$
  register, Strang steps alternating the wave circuit
  $\mathbf{W}(\tau)\approx e^{-i\tau \mathbf{H}}$ (system register only) with the damping
  phases $e^{+i\tau \mathbf{D}_\eta\otimes\boldsymbol{\Sigma}}$,
  closing $\QFT$, and post-selection of the most significant one or two
  $p$ qubits.  (b)~One factor of $\mathbf{W}(\tau)$: the
  Bell-basis evolution $e^{-i\tau(\zeta Q+\bar\zeta Q^{\dagger})}$ of the term
  $Q=\sigma_{01}\otimes\sigma_{00}\otimes I\otimes I\otimes
  \sigma_{01}\otimes\sigma_{10}$ ($\phi=\arg\zeta$) of the 2D acoustic Hamiltonian on a
  $4\times4$ grid ($q_5,q_4$: block label; $q_3,q_2$: $x$; $q_1,q_0$:
  $y$).  CNOTs rotate the ladder qubits to the Bell basis, and the single
  multi-controlled $R_z$ carries the block projector $\sigma_{00}$
  ($q_4$) as an open-circle $0$-control.  $\mathbf{W}(\tau)$ is the palindromic
  product of $4n_x+4$ such factors, each at half step.}
  \label{fig:circuits}
\end{figure}
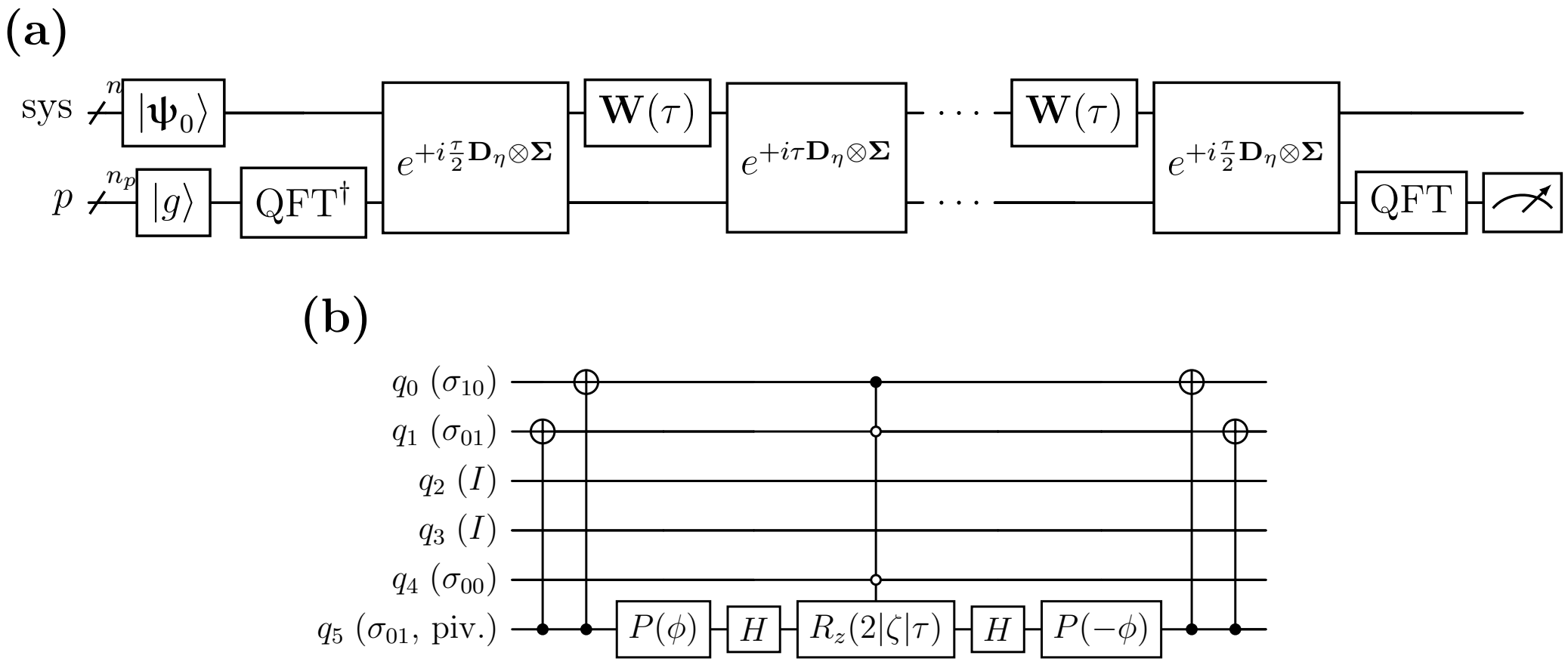

Step 3 of the algorithm, $e^{-iT\hat{\boldsymbol{\mathcal H}}}$ with
$\hat{\boldsymbol{\mathcal H}}=\mathbf{I}\otimes \mathbf{H}-\mathbf{D}_\eta\otimes\boldsymbol{\Sigma}$, is the only nontrivial
one.  We Trotterize it~\cite{trotter1959,suzuki1991} into a wave part, $e^{-i\tau \mathbf{I}\otimes \mathbf{H}}$, which
acts on the system register alone, and a damping part,
$e^{+i\tau \mathbf{D}_\eta\otimes\boldsymbol{\Sigma}}$, which is diagonal on the whole
register.  We compile both into elementary gates.

\subsection{Difference operators as qubit strings}
\label{sec:strings}

We use the single-qubit operators
$\sigma_{01}=\ket0\!\bra1$, $\sigma_{10}=\ket1\!\bra0$ and the projectors
$\sigma_{00}=\ket0\!\bra0$, $\sigma_{11}=\ket1\!\bra1$; in a tensor
product of such operators (a string) the leftmost factor acts on the
most significant qubit.  The non-periodic
shift on $n$ qubits is a sum of $n$ ladder strings~\cite{sato2024hamiltonian},
\begin{equation}
  \mathbf{S}=\sum_{k=1}^{n} I^{\otimes(n-k)}\otimes\sigma_{01}\otimes
    \sigma_{10}^{\otimes(k-1)},
  \label{eq:shift}
\end{equation}
the $k$th string implementing the carry $\ket{j+1}\to\ket j$ in which the
lowest $k-1$ bits of $j+1$ are zero (e.g.\ $I\otimes I\otimes\sigma_{01}$,
$I\otimes\sigma_{01}\otimes\sigma_{10}$ and
$\sigma_{01}\otimes\sigma_{10}\otimes\sigma_{10}$ for $n=3$).  Hence $\Df=(\mathbf{S}-\mathbf{I})/h$ and $\Db=(\mathbf{I}-\mathbf{S}^{\dagger})/h$
are sums of $n+1$ strings each, and the coupling of block $b$ to block
$b'$ in $\mathbf{H}$ is $\ket b\!\bra{b'}=\bigotimes_i\sigma_{b_ib'_i}$,
one factor per label qubit, tensored with the spatial string.  For the
2D acoustic system the couplings are
$\ket{00}\!\bra{01}=\sigma_{00}\otimes\sigma_{01}$ and
$\ket{00}\!\bra{10}=\sigma_{01}\otimes\sigma_{00}$ with their Hermitian
partners.  Counting strings per spatial direction gives the term counts
\begin{equation}
  \begin{aligned}
  \text{acoustic: }&\ 2\textstyle\sum_a (n_a+1),\\
  \text{elastic: }&\ 2\,\nu_d\textstyle\sum_a (n_a+1),
  \end{aligned}
  \label{eq:termcount}
\end{equation}
where $\nu_d$ is the number of velocity--stress couplings per direction
($\nu_1=2$ for P--SV, $\nu_2=3$, $\nu_3=5$; e.g.\ in 2D the $x$ couplings are
$s_{xx}\!\leftrightarrow\!\tilde v_x$, $s_{yy}\!\leftrightarrow\!\tilde v_x$
and $s_{xy}\!\leftrightarrow\!\tilde v_y$).  For 2D elastic this is
$6n_x+6n_y+12=6n_{\rm sys}-6$, which we confirm numerically.
All counts are linear in the number of qubits.  \ref{app:strings}
gives the derivation.

\subsection{Bell-basis term evolution with projectors}
\label{sec:bell}

Each Hermitian pair $\zeta Q+\bar\zeta Q^{\dagger}$, with $Q$ a string over
$\{I,\sigma_{01},\sigma_{10},\sigma_{00},\sigma_{11}\}$ that contains at
least one ladder operator, couples exactly two basis states inside every
subspace selected by its projectors and identities.  A ladder of
CNOTs from a pivot qubit rotates the pair into a Bell basis
in which the term is $|\zeta|\,Z$ on the pivot, conditioned on the other
ladder qubits.  The evolution is then one multi-controlled
$R_z(2|\zeta|\tau)$ between phase gates $P(\pm\arg\zeta)$ and Hadamards, and
the CNOT ladder is undone afterwards~\cite{sato2024hamiltonian}.
Block projectors $\sigma_{00}$ and $\sigma_{11}$ enter as plain $0$- and
$1$-controls of that rotation, which confines the rotation to the
selected block pair.  Figure~\ref{fig:circuits}(b) shows the compiled
term $Q=\sigma_{01}\otimes\sigma_{00}\otimes I\otimes I\otimes
\sigma_{01}\otimes\sigma_{10}$ of the 2D acoustic Hamiltonian.  This projector
rule is what makes separated multi-field systems compilable: the block
couplings of Eqs.~\eqref{eq:Hdisc} and \eqref{eq:Helasticdisc} are exactly
such strings.  The rule is exact: a first-order product of the term exponentials
differs from $e^{-i\tau \mathbf{H}}$ only by the Trotter error, $O(\tau^{2})$
per step.  The product is symmetrized (palindromic order of the terms), giving a
second-order approximation $\mathbf{W}(\tau)=e^{-i\tau \mathbf{H}}+O(\tau^{3})$ at twice
the gate count.

\subsection{Trotterization}
\label{sec:trotter}

With $\tau=T/n_t$ the outer Strang splitting~\cite{strang1968} reads
\begin{equation}
  e^{-i\tau\hat{\boldsymbol{\mathcal H}}}
  \approx e^{+i\frac{\tau}{2}\mathbf{D}_\eta\otimes\boldsymbol{\Sigma}}
  \bigl(\mathbf{I}\otimes \mathbf{W}(\tau)\bigr)\,
  e^{+i\frac{\tau}{2}\mathbf{D}_\eta\otimes\boldsymbol{\Sigma}},
  \label{eq:strang}
\end{equation}
and consecutive half-steps of damping merge, so $n_t$ steps cost $n_t$
wave circuits and $n_t+1$ damping factors
[Fig.~\ref{fig:circuits}(a)].  The local error is
$O(\tau^{3})$ with constants given by the nested commutators of the two
parts~\cite{childs2021trotter},
$[\mathbf{D}_\eta\otimes\boldsymbol{\Sigma},[\mathbf{D}_\eta\otimes\boldsymbol{\Sigma},\mathbf{I}\otimes \mathbf{H}]]
=\mathbf{D}_\eta^{2}\otimes[\boldsymbol{\Sigma},[\boldsymbol{\Sigma},\mathbf{H}]]$ and
$[\mathbf{I}\otimes \mathbf{H},[\mathbf{I}\otimes \mathbf{H},\mathbf{D}_\eta\otimes\boldsymbol{\Sigma}]]=\mathbf{D}_\eta\otimes[\mathbf{H},[\mathbf{H},\boldsymbol{\Sigma}]]$.
The basic commutator has entries
\begin{equation}
  [\boldsymbol{\Sigma},\mathbf{H}]_{rs}=(\sigma_r-\sigma_s)\,H_{rs},
  \label{eq:commutator}
\end{equation}
which are nonzero only between neighbouring grid values whose damping
rates differ, i.e.\ inside the graded layer.  There
$|\sigma_r-\sigma_s|\lesssim m\,\sigma_{\max}/\nL$, so
$\|[\boldsymbol{\Sigma},\mathbf{H}]\|\lesssim\|\mathbf{H}\|\,m\sigma_{\max}/\nL$: the splitting
error is controlled by the gradient $\sigma_{\max}/\nL$ of the
profile, and a wider, gentler layer is also easier to Trotterize.  The
commutators also carry powers of $\mathbf{D}_\eta$, whose norm $\pi/\Delta p$
grows with the $p$ resolution; the high-$|\eta|$ modes, however, carry
little amplitude, because $\hat g(\eta)$ decays as $|\eta|^{-3}$ for the
$C^{1}$ profile.  The measured error constants are modest.  For the 2D
acoustic sponge ($8\times8$, 17 qubits, $T=10$, $n_p=9$) the error
falls as $4.1\times10^{-2}\to1.0\times10^{-2}\to2.7\times10^{-3}$ for
$n_t=50/100/200$, with fitted orders $-0.97$ (first-order splitting) and
$-2.01$ (Strang).  For a $4\times4$ elastic sponge ($c_s=0.5$, $n_p=5$)
they are $-1.003/-2.001$.

\subsection{A polynomial-size damping circuit}
\label{sec:structured}

The damping factor $\mathbf{U}(\tau)=e^{+i\tau \mathbf{D}_\eta\otimes\boldsymbol{\Sigma}}$ is diagonal,
with entries $e^{i\tau\eta_k\sigma_s}$ on $\ket k\otimes\ket s$.  A
generic diagonal unitary on $n=n_p+n_{\rm sys}$ qubits compiles into
$\Theta(2^{n})$ gates~\cite{bullock2004diagonal}; with Qiskit's synthesis
we measure exactly $2^{n}-2$ CNOTs at every size tested
($n=6,\dots,14$ and $n=15,17,19,21$).  This is
exponential in the number of qubits, and at 17--26 qubits it exceeds the
wave step by two to four orders of magnitude.
The structure of $\eta$ and of $\boldsymbol{\Sigma}$ removes this cost.

First, $\eta$ is linear in the $p$ bits.  Writing the index as
$k=\sum_{b=0}^{n_p-1}2^{b}k_b$, its two's-complement reading in
Eq.~\eqref{eq:eta} is
\begin{equation}
\begin{split}
  \bar k&=\sum_{b=0}^{n_p-2}2^{b}k_b-2^{n_p-1}k_{n_p-1}
  =:\sum_{b}\chi_b\,k_b,\\
  \eta_k&=\frac{\pi}{p_{\max}}\sum_b\chi_bk_b .
\end{split}
  \label{eq:etalinear}
\end{equation}
Therefore
\begin{equation}
\mathbf{U}(\tau)=\prod_{b}\Bigl[\ket0\!\bra0_b\otimes\mathbf{I}
+\ket1\!\bra1_b\otimes e^{i\tau(\pi/p_{\max})\chi_b\boldsymbol{\Sigma}}\Bigr]:
\end{equation}
$n_p$ commuting factors, each a system-register diagonal controlled by a
single $p$ qubit.

Second, $\boldsymbol{\Sigma}$ is a sum of one-dimensional profiles.  By
Eq.~\eqref{eq:Sigmadisc}, on every block $\boldsymbol{\Sigma}$ is a sum
$\sum_a f_a(i_a)$ of profiles that each depend on one coordinate index.
Hence $e^{i\theta\boldsymbol{\Sigma}}$ factorizes over blocks and axes.

Third, each profile is a low-degree polynomial on dyadic strips.
The polynomial profile of Eq.~\eqref{eq:profile} sampled at
$x_j=(j+\text{stagger})h$ is a polynomial of degree $m$ in the index $j$
on each layer.  When $\nL$ is a power of two, each layer is a
dyadic interval: the left layer $j<\nL$ is the set where the
high $n_a-\log_2\nL$ bits of $j$ are all $0$, and the right layer is the
set where they are all $1$ (up to one node for staggered sampling).  A
polynomial of degree $m$ in
$j=\sum_c2^{c}j_c$ is, because $j_c^{2}=j_c$, a multilinear polynomial
of degree at most $m$ in the bits:
\begin{equation}
\begin{split}
  f(j)&=\sum_{|S|\le m}a_S\prod_{c\in S}j_c,\\
  a_S&=\sum_{R\subseteq S}(-1)^{|S|-|R|}f\Bigl(\sum_{c\in R}2^{c}\Bigr),
\end{split}
  \label{eq:mobius}
\end{equation}
the coefficients being the M\"obius transform of the values.  For
example, on the left strip of a $\nL=4$, $m=2$ layer,
$(4-j)^{2}=16-7j_0-12j_1+4j_0j_1$.  Single nodes where the sampled
profile deviates from the polynomial (the clipping at the outer edge
and the extra node $j=\nL$ of a staggered field) are pieces of their
own.  An adaptive bisection finds
the decomposition of an arbitrary profile into dyadic pieces on
which Eq.~\eqref{eq:mobius} is exact.

\begin{theorem}[Structured damping circuit]
\label{thm:structured}
Let $\boldsymbol{\Sigma}=\sum_{\beta}\ket\beta\!\bra\beta\otimes\sum_af^{(\beta)}_a$
with each one-dimensional profile a sum of $K$ dyadic pieces of
size $2^{q}$ on which it is a polynomial of degree $\le m$.  Then
\begin{equation}
\begin{split}
  e^{+i\tau \mathbf{D}_\eta\otimes\boldsymbol{\Sigma}}
  =\prod_{b=0}^{n_p-1}\;\prod_{\beta,a}\;\prod_{\ell}\;
   \prod_{|S|\le m}
   \mathrm{C}^{\ell,\beta,S}\!P_{k_b}&\\
   \Bigl(\tau\frac{\pi\,\chi_b}{p_{\max}}
   a^{(\ell)}_S\Bigr)&
\end{split}
  \label{eq:structured}
\end{equation}
exactly, where $\ell$ runs over the dyadic pieces and
$\mathrm C^{\ell,\beta,S}P_{k_b}(\theta)$ is a phase
$e^{i\theta}$ on $p$ qubit $b$, controlled by the block-label bits of
$\beta$, the fixed high bits of the dyadic piece $\ell$ (as $0$- or
$1$-controls), and the bits in $S$.  All factors commute.  The number
of gates is at most
$n_p\,\#\{(\beta,a)\}\,K\sum_{r\le m}\binom{q}{r}=O(n_p\,K\,q^{m})$, each
with at most $n_{\rm lab}+n_a$ controls.  With one clean flag ancilla
that holds the piece predicate (computed and uncomputed by two
multi-controlled NOTs per piece), every phase has at most $m+1$
controls.  \ref{app:structured} details the construction.
\end{theorem}

\begin{proof}
The phase on $\ket k\otimes\ket s$ is
\begin{equation*}
\tau\eta_k\sigma_s=\tau\frac{\pi}{p_{\max}}\sum_b\chi_bk_b
\sum_{\beta,a,\ell}\mathbb 1_{\beta,\ell}(s)\sum_Sa^{(\ell)}_S\prod_{c\in
S}j_c ,
\end{equation*}
a sum of products of bits.  The exponential of a sum of
commuting diagonal terms is the product of their exponentials, and
$e^{i\theta x_1\cdots x_r}$ is the phase gate on one of the bits
controlled by the others.
\end{proof}

For a layer whose width is a fixed fraction of the grid, $K=O(1)$
pieces per profile and $q=n_a-O(1)$, so the damping circuit has
$O(n_p\,n_a^{m})$ phase gates, polylogarithmic in the number of grid
points.  For $m=2$ and ancilla-free synthesis of the multi-controlled
phases (quadratic in the control count) this is
$O(n_p\,n_a^{2}\,n_{\rm sys}^{2})$ CNOTs; with the flag ancilla
it is $O(n_p\,n_a^{2}+K n_{\rm sys}^{2})$.  The construction is exact
for any profile and any stagger; only the gate count depends on the
structure.  Combined with the $O(n^{2})$ wave step, the complete
Trotter step of the Schr\"odingerized sponge is polynomial in the number
of qubits.  We verify Eq.~\eqref{eq:structured} against the dense
diagonal to $10^{-10}$ in operator norm for acoustic and elastic
sponges, with and without the flag ancilla, and end to end in
the gate-level runs.

\subsection{State preparation and total cost}
\label{sec:cost}

The initial state is a product, $\ket g\otimes\ket{\boldsymbol{\uppsi}_0}$.  The
profile state $\ket g$ lives on the $n_p$-qubit $p$ register, whose size
does not depend on the grid.  A generic preparation costs $O(2^{n_p})$
gates, a constant with respect to the spatial resolution.  For the
kinked profile $e^{-|p|}$ it is $O(n_p)$: on each half of the $p$ grid
$e^{-|p_j|}$ is the exponential of a linear function of $j$, hence a
product over the bits, and one rotation of the most significant qubit
followed by controlled single-qubit rotations prepares it.  The initial
wavefield $\boldsymbol{\uppsi}_0$ is problem-specific; smooth separable data such as
our Gaussian pulses are efficiently preparable~\cite{grover2002}.  In
our demonstrations we load the product
state by exact state preparation.

With $n_t=T/\tau$ steps and second-order splitting,
$\tau=O\bigl(\sqrt{\varepsilon/(C_2T)}\bigr)$ for a target splitting
error $\varepsilon$, and the total gate count is
\begin{equation}
\begin{split}
  G_{\rm total}=\;&O\Bigl(\frac{T^{3/2}}{\sqrt\varepsilon}\,
  \bigl[G_{\rm wave}(n_{\rm sys})+G_{\rm damp}(n_p,n_{\rm sys})\bigr]\Bigr)\\
  &+2\,G_{\QFT}(n_p),
\end{split}
  \label{eq:totalcost}
\end{equation}
both bracketed terms polynomial in $n_{\rm sys}$, times
$O(\|\boldsymbol{\uppsi}_0\|/\|\boldsymbol{\uppsi}(T)\|)$ repetitions for the post-selection.  For the
2D acoustic system the wave step fits
$G_{\rm wave}(n)=16.7n^{2}-112.7n+293$ elementary operations in the
\{CX, $u$\} basis.  The commutator constant $C_2$ depends on $\Delta p$
through $\mathbf{D}_\eta$, a dependence we have only
measured, not bounded.

%% file: sections/numerics.tex
\section{Numerical results}
\label{sec:numerics}

\begin{figure}[tbp]
  \centering
  \includegraphics[width=\textwidth]{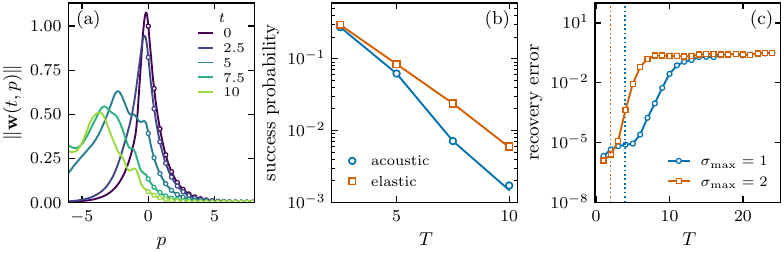}
  \caption{The warped variable.  (a)~$\|\mathbf{w}(t,p)\|$ for the $8\times8$
  acoustic sponge (lines) and $e^{-p}\|\boldsymbol{\uppsi}(t)\|$ (dots): the two agree
  on $p\ge0$ (Proposition~\ref{prop:exact}; up to $t\le T_{\rm dep}$ on
  the whole window), while the
  absorbed part is transported to $p<0$.  (b)~Probability that the most
  significant $p$ qubit reads $1$ ($p\ge0$), from the exactly evolved
  Schr\"odingerized state (symbols) and from Eq.~\eqref{eq:prob} (lines), acoustic and elastic
  $8\times8$ sponges; it decays with the energy remaining in the domain.
  (c)~Absolute recovery error $\|\mathbf{z}_{\rm rec}-\boldsymbol{\uppsi}(T)\|/\|\boldsymbol{\uppsi}_0\|$
  versus $T$ at fixed
  $p_{\max}=8$, $n_p=9$, for $\sigma_{\max}=1$ and $2$ (corner rate
  $\sbar=2\sigma_{\max}$).  Dotted lines: the domain-of-dependence time
  $T_{\rm dep}=(p_{\max}-p^{*})/\sbar$ of Proposition~\ref{prop:dod},
  up to which the error is at the $p$-discretization floor.}
  \label{fig:pspace}
\end{figure}

\begin{figure}[tbp]
  \centering
  \includegraphics[width=\textwidth]{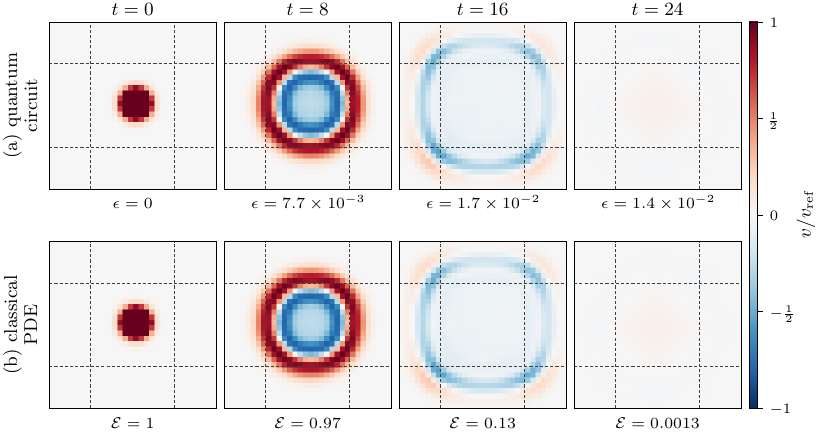}
  \caption{Gate-level absorbed acoustic wave: $32\times32$ grid, 21
  qubits (12 system $+$ 9 $p$), up to 240 second-order Trotter steps.
  Top: field recovered from the quantum circuit at $t=0,8,16,24$
  (dashed: layer strips).  Bottom: classical solution of the damped PDE
  (finite differences in space, exact in time).  $\epsilon$: relative error of the
  recovered field; $\mathcal E=\|\boldsymbol{\uppsi}(t)\|^{2}/\|\boldsymbol{\uppsi}_0\|^{2}$: energy
  remaining in the domain.  This run is also shown in
  Ref.~\cite{nguyen2026companion}.}
  \label{fig:gate21}
\end{figure}

\begin{figure}[tbp]
  \centering
  \includegraphics[width=\textwidth]{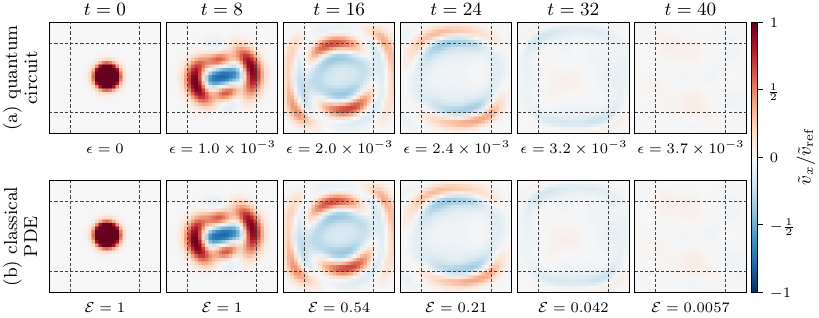}
  \caption{Gate-level absorbed elastic wave: $32\times32$ grid, 25
  qubits (13 system $+$ 11 $p$ $+$ 1 flag), structured damping circuit,
  snapshots of one 1280-step run.  (a)~$\tilde v_x$ recovered from the
  quantum circuit; (b)~the classical solution of the damped PDE
  (finite differences in space, exact in time), on one colour scale;
  dashed lines mark the layers.  $\epsilon$ and $\mathcal E$ as in
  Fig.~\ref{fig:gate21}.}
  \label{fig:elastic}
\end{figure}

All runs use $h=1$, the $C^{1}$ profile of Eq.~\eqref{eq:gcubic}, Gaussian
initial pulses, $c=1$ for the acoustic system and $c_p=1$,
$c_s=1/\sqrt3$, $\rho=1$ for the elastic one.  Statevector simulations
use Qiskit~1.4.5 and Aer~0.16.4~\cite{qiskit2024}.

\subsection{The warped variable in $p$ space}
\label{sec:pspace}

Figure~\ref{fig:pspace}(a) shows $\|\mathbf{w}(t,p)\|$ for the $8\times8$
acoustic sponge ($\nL=2$, $\sigma_{\max}=1$, $n_p=10$, $p_{\max}=16$),
computed by exact per-mode evolution of Eq.~\eqref{eq:Hhat}.  On $p\ge0$
it coincides with $e^{-p}\|\boldsymbol{\uppsi}(t)\|$ (dots), as
Proposition~\ref{prop:exact} requires; the domain-of-dependence time is
$T_{\rm dep}=(16-p^{*})/2\approx7.95$, so at $t=10$ the top of the
window has begun to receive seam data.  The mass that has left the
physical solution has been transported to $p<0$, and the profile on
$p<0$ moves left as the sponge absorbs; this is the
multi-mode version of Example~\ref{ex:scalar} and the analogue of
Fig.~5 of \citet{jin2023schrodingerisation}.  The recovered field
has relative errors of $7\times10^{-6}$, $2.2\times10^{-5}$ and
$8.6\times10^{-5}$ at $t=2.5,5,7.5$, the $p$-discretization error alone,
and $2.0\times10^{-4}$ at $t=10$, where the slice near $p=0$ is still
uncontaminated.

Figure~\ref{fig:pspace}(b) compares the
probability of the outcome ``most significant $p$ qubit $=1$'' with
Eq.~\eqref{eq:prob}.  At $T=0$ both equal $\kappa_g=0.408$.  For
$T=2.5$--$7.5$ they agree to all printed digits
($0.276$, $0.0623$, $7.14\times10^{-3}$ acoustic;
$0.297$, $0.0843$, $0.0239$ elastic) while the remaining energy falls by
two orders of magnitude.  At $T=10$ they separate by $15\%$
(acoustic) and $2.5\%$ (elastic): the top of the $p\ge0$ window has
started to receive seam data.  The post-selected system state has
infidelity $\le6.7\times10^{-6}$ with $\boldsymbol{\uppsi}(T)/\|\boldsymbol{\uppsi}(T)\|$ up to
$T=7.5$, and $0.13$ (acoustic) and $0.025$ (elastic) at $T=10$.  The
two-qubit post-selection $p\in[0,p_{\max}/2)$ keeps the infidelity at
$3.8\times10^{-7}$ (acoustic) and $2.4\times10^{-8}$ (elastic) at
$T=10$.  It should be the default.

Figure~\ref{fig:pspace}(c) fixes
$p_{\max}=8$ and increases $T$.  Up to the domain-of-dependence time
$T_{\rm dep}$ of Proposition~\ref{prop:dod} ($4.0$ and $2.0$ for
$\sigma_{\max}=1$ and $2$) the absolute error stays at the
discretization floor ($\le8\times10^{-6}$ and $\le2.5\times10^{-6}$ of
$\|\boldsymbol{\uppsi}_0\|$).  It remains small somewhat beyond $T_{\rm dep}$
($2.5\times10^{-5}$ at $T=6$ for $\sigma_{\max}=1$), because only the energy that actually sits in the
corners of the layer moves at the worst-case speed $\sbar$.  It then
grows and saturates at $\sim0.3\|\boldsymbol{\uppsi}_0\|$ once the absorbed mass has
wrapped around.  On this tiny grid much of the energy sits in the layer
at all times, the least favourable case.  On larger grids the
rule of Eq.~\eqref{eq:pmaxrule} is very conservative: a wave that has crossed
the layer and returned has accumulated a shift of only about
$\ln(1/R_0)\,\sigma_{\max}/\sigma_0\approx7$ in $p$ (the attenuation
exponent), not $\sbar T$.  The $32\times32$
acoustic run below uses $p_{\max}=18$ although $p^{*}+\sbar T\approx62$,
and it recovers to $1.7\times10^{-2}$.

\subsection{One-dimensional gate-level run}
\label{sec:onedim}

The one-dimensional gate-level run uses a 64-node grid ($n_x=6$,
7 system qubits, Gaussian pulse of width $2.5h$) with
$n_p=9$, $p_{\max}=16$, $\nL=8$, $\sigma_{\max}=\sigma_0$, the
structured damping circuit with one flag ancilla (17 qubits, 3798
CNOTs per step for the damping) and $\tau=0.05$.  The recovered state
matches $e^{\mathbf{A}T}\boldsymbol{\uppsi}_0$ to $1.1\times10^{-3}$, $2.2\times10^{-3}$ and
$5.1\times10^{-3}$ at $T=10$, $20$ and $30$ (200--600 steps); by $T=30$
the pulse has entered the layers and $85\%$ of its energy has been
absorbed.

\subsection{Gate-level simulations}
\label{sec:gatelevel}

Figure~\ref{fig:gate21}
shows the gate-level Schr\"odingerized sponge on a $32\times32$ grid
(12 system $+$ 9 $p$ qubits, $\nL=8$, $\sigma_{\max}=\sigma_0$, Gaussian pulse of
width $1.5h$, $p_{\max}=18$, $\tau=0.1$, up to
240 Strang steps).  The wave part is the Bell-basis circuit (983
native simulator operations, or 1343 in the \{CX, $u$\} basis, per
step); the damping part in this run is a dense diagonal gate.  The recovered field matches the exact
$e^{\mathbf{A}T}\boldsymbol{\uppsi}_0$ to $7.7\times10^{-3}$, $1.7\times10^{-2}$ and
$1.4\times10^{-2}$ at $T=8,16,24$, while the energy in the domain falls
from $0.97$ to $1.25\times10^{-3}$.  The outgoing ring is absorbed in the
layer with no visible reflection.

Figure~\ref{fig:elastic} shows the gate-level elastic sponge on a
$32\times32$ grid (13 system $+$ 11 $p$ qubits $+$ 1 flag, $\nL=6$,
$\sigma_{\max}=0.5$, Gaussian pulse of width $2h$, $p_{\max}=40$,
$\Delta p=0.039$, $\tau=0.05$), the largest run in this paper: a single
time march of 1280 Strang steps to $T=64$, with the damping compiled by
the structured circuit of Theorem~\ref{thm:structured} (8448 CNOTs per
step) and snapshots taken every 160 steps.  The P
and S fronts separate, enter the layers and leave; the quantum and
classical fields are indistinguishable at plotting resolution.  The
relative errors are $1.0\times10^{-3}$, $2.0\times10^{-3}$,
$2.4\times10^{-3}$, $3.2\times10^{-3}$ and $3.7\times10^{-3}$ at
$T=8,16,24,32,40$, and stay at $3.6$--$3.9\times10^{-3}$ up to $T=64$,
while the energy in the domain falls to $7.2\times10^{-4}$.  The error
does not grow once the wave has left: the late-time Trotter error is
carried by an ever smaller field.  The two-qubit window estimator agrees
with the single slice to within $0.5\%$ at every time.  The same setup on
a $16\times16$ grid (23 qubits, $\nL=3$, $\sigma_{\max}=1$, 800 steps to
$T=40$) gives $1.7$--$3.0\times10^{-3}$, and on $8\times8$ (19 qubits)
$4.9\times10^{-4}$--$1.3\times10^{-3}$ for $T=2$--$8$
(Table~\ref{tab:structured}).

We repeated the
$8\times8$ acoustic and elastic runs with the damping factor compiled by
Theorem~\ref{thm:structured} (flag-ancilla variant).  The time loop then
contains only the Bell wave circuit and multi-controlled phase and
NOT gates; no dense diagonal remains.  Table~\ref{tab:structured}
lists the results.  The flag ancilla returns to $\ket0$ exactly (leakage
$0$ to machine precision).  The errors coincide with those of the
dense-diagonal runs (e.g.\ elastic $4.85\times10^{-4}$,
$8.52\times10^{-4}$, $1.16\times10^{-3}$, $1.31\times10^{-3}$, identical
to three digits), as they must, since the two circuits are equal as
unitaries.  The window estimator obtained by two-qubit post-selection,
$\hat{\boldsymbol{\uppsi}}=\sum_{j\in J}e^{-p_j}\mathbf{w}_j/\sum_{j\in J}e^{-2p_j}$, is about
as accurate as the single slice (within $30\%$).  Each step costs 218
(acoustic) or 1150 (elastic) CNOTs for the wave and 3376 or 2936 for the
damping (the acoustic count differs from Table~\ref{tab:gatecounts}
because this run uses $\nL=2$).

\begin{table}[tbp]
  \caption{Gate-level runs with the structured damping circuit
  ($8\times8$ grids, $\tau=0.05$, $n_p=9$; acoustic $p_{\max}=16$,
  elastic $p_{\max}=10$; one flag ancilla).  $\mathcal E$: energy remaining,
  $\|\boldsymbol{\uppsi}(T)\|^{2}/\|\boldsymbol{\uppsi}_0\|^{2}$; $\epsilon_{\rm slice}$, $\epsilon_{\rm win}$: relative error of the
  state recovered from a single slice and from the two-qubit window; $\Pr$: probability of the
  window outcome.}
  \label{tab:structured}
  \centering
  \begin{tabular}{lcccc}
    \toprule
    $T$ (steps) & $\mathcal E$ & $\epsilon_{\rm slice}$ & $\epsilon_{\rm win}$ & $\Pr$\\
    \midrule
    \multicolumn{5}{l}{acoustic, 18 qubits}\\
    2.5 (50)  & 0.677 & $8.4\times10^{-4}$ & $8.3\times10^{-4}$ & 0.285\\
    5.0 (100) & 0.153 & $1.7\times10^{-3}$ & $1.8\times10^{-3}$ & 0.064\\
    7.5 (150) & 0.018 & $3.0\times10^{-3}$ & $2.9\times10^{-3}$ & 0.0074\\
    10 (200)  & 0.004 & $2.7\times10^{-3}$ & $2.5\times10^{-3}$ & 0.0015\\
    \multicolumn{5}{l}{elastic, 19 qubits}\\
    2 (40)    & 0.833 & $4.9\times10^{-4}$ & $4.9\times10^{-4}$ & 0.342\\
    4 (80)    & 0.372 & $8.5\times10^{-4}$ & $8.5\times10^{-4}$ & 0.153\\
    6 (120)   & 0.123 & $1.2\times10^{-3}$ & $1.2\times10^{-3}$ & 0.051\\
    8 (160)   & 0.045 & $1.3\times10^{-3}$ & $1.7\times10^{-3}$ & 0.018\\
  \bottomrule
  \end{tabular}
\end{table}

\subsection{Gate counts}
\label{sec:gatecount}

\begin{figure}[tbp]
  \centering
  \includegraphics[width=\textwidth]{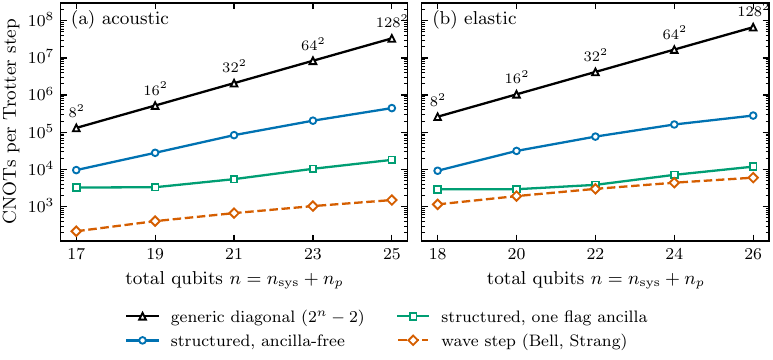}
  \caption{CNOTs per Trotter step versus total qubit count
  ($n_p=9$; grid size annotated): damping factor as a generic diagonal
  ($2^{n}-2$), as the structured circuit of
  Theorem~\ref{thm:structured} without and with one flag ancilla, and
  the Bell wave step.}
  \label{fig:gatecounts}
\end{figure}

Figure~\ref{fig:gatecounts} and Table~\ref{tab:gatecounts} compare, per
Trotter step and after transpilation to \{CX, $u$\}, the
damping factor compiled as a dense diagonal ($2^{n}-2$ CNOTs),
the structured circuit with and without the flag ancilla, and the wave
circuit.  Grids run from $8\times8$ to $128\times128$ with $\nL=N/8$
and $n_p=9$.  The dense diagonal doubles with every qubit and exceeds
the wave step by a factor of $600$ (acoustic, $8\times8$) to
$2\times10^{4}$ (acoustic, $128\times128$).  The structured circuit
grows polynomially.  With the flag ancilla it needs $2.9\times10^{3}$
to $1.8\times10^{4}$ CNOTs over the whole range, which is
$40$--$1850$ times (acoustic) and $89$--$5600$ times (elastic) fewer
than the dense diagonal.  For the elastic system at $128\times128$ the
damping is within a factor of two of the wave step.

\begin{table}[tbp]
  \caption{CNOT counts per Trotter step ($n_p=9$, $\nL=N/8$,
  $\tau=0.1$; transpilation to \{CX, $u$\} at optimization level~1).
  ``generic'': $2^{n}-2$, the count measured at every size tested up to
  $n=21$ (Sec.~\ref{sec:structured}) and used as a formula for the other rows; ``struct.'': ancilla-free structured circuit, whole-circuit
  transpilation; ``+flag'': with one extra ancilla qubit, sum of per-gate
  syntheses, each verified exact (\ref{app:structured}).}
  \label{tab:gatecounts}
  \centering
  \begin{tabular}{lrrrrr}
    \toprule
    grid & $n$ & generic & struct. & +flag & wave\\
    \midrule
    \multicolumn{6}{l}{acoustic}\\
    $8^{2}$   & 17 & 131070   & 9630   & 3268  & 218\\
    $16^{2}$  & 19 & 524286   & 28080  & 3332  & 408\\
    $32^{2}$  & 21 & 2097150  & 83808  & 5496  & 670\\
    $64^{2}$  & 23 & 8388606  & 205344 & 10428 & 1036\\
    $128^{2}$ & 25 & 33554430 & 447336 & 18140 & 1506\\
    \multicolumn{6}{l}{elastic}\\
    $8^{2}$   & 18 & 262142   & 9252   & 2936  & 1150\\
    $16^{2}$  & 20 & 1048574  & 31536  & 2944  & 1924\\
    $32^{2}$  & 22 & 4194302  & 76608  & 3856  & 3010\\
    $64^{2}$  & 24 & 16777214 & 162144 & 7168  & 4408\\
    $128^{2}$ & 26 & 67108862 & 281952 & 11920 & 6022\\
  \bottomrule
  \end{tabular}
\end{table}

%% file: sections/discussion.tex
\section{Discussion}
\label{sec:discussion}

The sponge damps every wave field at its own grid point.  The Hermitian
part of the generator, $\mathbf{H}_1=-\boldsymbol{\Sigma}$, is therefore
diagonal and negative semidefinite, and this one property simplifies
each step of the Schr\"odingerization pipeline.

First, the recovery has no threshold.  For a general generator the
solution can be read from the auxiliary variable only at
$p^{*}\ge\lamax^{+}(\mathbf{H}_1)\,T$, a threshold that grows with the
simulated time whenever $\mathbf{H}_1$ has a positive eigenvalue, that
is, whenever the dynamics can amplify some component.  The sponge only
removes energy, so $\lamax^{+}(\mathbf{H}_1)=0$ and
$\mathbf{w}(t,p)=e^{-p}\boldsymbol{\uppsi}(t)$ holds on the whole
half-line $p\ge0$ at every time (Proposition~\ref{prop:exact}).

Second, the truncation of the auxiliary axis has an explicit sufficient
condition.  In the variable $p$ the damping becomes transport
[Eq.~\eqref{eq:warpedpde}]: each field component moves to the left at
its own local damping rate, and none moves faster than the largest rate
$\sbar$.  On a finite periodic interval $|p|<p_{\max}$ the only error is
the data that enters through the end of the interval, and it needs the
time $(p_{\max}-p^{*})/\sbar$ to reach the slice $p^{*}$ where the
solution is read.  The truncation is therefore exact, not merely
accurate, as long as $p_{\max}\ge p^{*}+\sbar T$
(Proposition~\ref{prop:dod}).  The condition contains only known
quantities, the largest damping rate and the simulated time, and
meeting it costs $\log_2(\sbar T)$ qubits.

Third, the measurement is simple and its cost is physical.  Every slice
$p\ge0$ carries the same state, so no single slice has to be selected:
measuring the one or two most significant qubits of the $p$ register is
enough.  The success probability is a constant $\kappa_g\approx0.4$
times the fraction of the energy that the sponge has not absorbed
(Proposition~\ref{prop:prob}), so the only loss beyond this constant is
the wave that has left the domain.

Fourth, the damping step is cheap.  Because $\boldsymbol{\Sigma}$ is
diagonal, the dissipative factor of each Trotter step is a diagonal
unitary; because the frequency $\eta$ is linear in the bits of the $p$
register and the damping profile is a low-degree polynomial on dyadic
strips, this unitary factorizes exactly into polynomially many
multi-controlled phases (Theorem~\ref{thm:structured}) instead of the
exponentially many gates of a generic diagonal.  The resulting algorithm
is fully explicit at the gate level and polynomial per time step.

The sponge is not a perfectly matched absorber~\cite{berenger1994pml}.
Its reflections are percent-level in amplitude on the grids we
simulated and decrease only slowly with the layer width.  On a quantum register wide layers are
cheap, since the qubit count grows with the logarithm of the grid size;
the gate count grows only through the term counts and the dyadic
structure of the profile.  The post-selection cost
$\|\boldsymbol{\uppsi}_0\|^{2}/\|\boldsymbol{\uppsi}(T)\|^{2}$ is the reciprocal of the energy that
remains; it is natural, since an absorbed wave has nothing left to
measure, but it means that late-time observables of a nearly empty
domain are expensive.

On a quantum register the
domain is cheap to enlarge: doubling it along an axis costs one qubit.
For a transient simulation with a known horizon $T$, padding each side
by about $cT/2$ keeps reflections out of the interior until time $T$ at
the price of roughly $\log_2(1+cT/L_{\rm d})$ extra qubits per axis ($L_{\rm d}$ is the
width of the unpadded domain), and the
padded evolution stays unitary, with no auxiliary register and no
post-selection.  (Reading out the interior costs the same either way: the
probability of finding the wave there equals its remaining energy
fraction.)  For short transients this is a valid alternative.  An
absorber is needed when the problem has no such horizon.  With a
continuous or long-running source, a closed domain never reaches a
steady state: energy accumulates and far-wall reflections always return
eventually, so time-harmonic (frequency-domain) responses and Green's
functions obtained by running to steady state require a radiation
condition, which only absorption supplies.  The same holds for
resonances and scattering, the reason complex absorbing potentials are
used in quantum dynamics.  Absorption also keeps the
domain fixed when $T$ is set by the physics, not chosen in advance, and
it truncates a velocity model given on a finite region.  Finally, real media dissipate
(intrinsic attenuation), so the wave dynamics is non-unitary whatever
the boundary.  The pipeline developed here (Schr\"odingerization of a
generator with $\mathbf{H}_1\preceq0$, structured damping circuits, and the
post-selection analysis) is the one such physics requires, and the
sponge is its cleanest instance.

Several extensions are immediate.  (i)~Three dimensions require no new idea:
Table~\ref{tab:qubits}, Eq.~\eqref{eq:termcount}, the sum
$\sigma=\sigma_x+\sigma_y+\sigma_z$ and Theorem~\ref{thm:structured}
cover it (\ref{app:3d}).  (ii)~Heterogeneous media enter
through variable coefficients in $\mathbf{H}$, e.g.\
$\mathbf{H}=\bigl[\begin{smallmatrix}0&\mathbf{B}\\\mathbf{B}^{\dagger}&0\end{smallmatrix}\bigr]$
with $\mathbf{B}=\diag(\rho^{-1/2})\,\mathbf{D}\,\diag(K^{1/2})$ ($K$ the bulk modulus)
for the acoustic system.  The sponge stays diagonal, so Lemma~\ref{lem:dissipative} and
Propositions~\ref{prop:exact}--\ref{prop:prob} are unchanged; only the
wave circuit acquires coefficient oracles~\cite{sato2025nonconservative}.
(iii)~Sources and receivers require
time-dependent inhomogeneous terms, which Schr\"odingerization handles
by augmentation~\cite{jin2024recovery}; with them, the solver becomes
the forward model of wave-based inversion such as full waveform
inversion~\cite{virieux2009overview}, which needs open-domain simulations by construction.  This
route is distinct from variational ones, in which parameterized quantum
circuits enter a physics-informed neural network that learns the
wavefield and the velocity model, as in a recent hybrid
quantum--classical approach to full waveform
inversion~\cite{nguyen2026hybrid}, and from quantum annealing
formulations of seismic inversion~\cite{nguyen2025annealing,vashisth2026quantum};
here the wavefield is evolved
by Hamiltonian simulation.  (iv)~On hardware the structured
damping circuit removes the last exponential component of the step.  The
remaining cost drivers are the multi-controlled rotations, whose
synthesis with ancillas is linear in the control count~\cite{barenco1995}.

\section{Conclusion}
\label{sec:conclusion}

We have derived and implemented, from the wave equation to elementary
gates, a quantum algorithm for acoustic and elastic waves in an open
domain.  The domain is opened by a dissipative sponge layer and the
dynamics is made unitary by Schr\"odingerization.  The sponge makes
every step of the algorithm explicit: recovery is exact for all
$p\ge0$, the auxiliary axis can be truncated at $p_{\max}\ge
p^{*}+\sbar T$, success is a single-qubit (or two-qubit) measurement
with probability equal to the unabsorbed energy times $\kappa_g\approx0.4$,
and the damping step is a polynomial-size circuit.  Gate-level
simulations on 17--25 qubits reproduce the classical solutions of the
damped acoustic and elastic wave equations with a relative energy-norm
error of $5\times10^{-4}$--$1.7\times10^{-2}$.

%% file: sections/appA.tex
\section{Three-dimensional formulas}
\label{app:3d}

For the acoustic system, with $\boldsymbol{\uppsi}=[\mathbf{v};\mathbf{q}_x;\mathbf{q}_y;\mathbf{q}_z]$ (labels $00,01,10,11$),
\begin{equation}
  \mathbf{H}=c\sum_{a\in\{x,y,z\}}\Bigl(\ket{0}\!\bra{\ell_a}\otimes\Df_a
     +\ket{\ell_a}\!\bra{0}\otimes(\Df_a)^{\dagger}\Bigr),
\end{equation}
$\ell_x=01$, $\ell_y=10$, $\ell_z=11$, so that the three couplings carry
the label factors $\sigma_{00}\otimes\sigma_{01}$,
$\sigma_{01}\otimes\sigma_{00}$ and $\sigma_{01}\otimes\sigma_{01}$.  $\Df_x=\Df\otimes
\mathbf{I}\otimes\mathbf{I}$, etc.  The sponge is
$\boldsymbol{\Sigma}=\diag(\boldsymbol{\Sigma}^{(v)},\boldsymbol{\Sigma}^{(q_x)},\boldsymbol{\Sigma}^{(q_y)},\boldsymbol{\Sigma}^{(q_z)})$
with $\sigma^{(v)}_{ijk}=\sigma_x(x_i)+\sigma_y(y_j)+\sigma_z(z_k)$ and
$q_a$ sampled half a cell back along axis $a$.  The register has
$n_x+n_y+n_z+2$ qubits and $\mathbf{H}$ has $2\sum_a(n_a+1)$ strings.

For the elastic system, with $\tilde{\mathbf{v}}=\sqrt\rho\,(v_x,v_y,v_z)$ and $\mathbf{s}=\tilde{\mathbf{C}}^{-1/2}
\tilde{\boldsymbol{\upsigma}}$, where
\begin{equation*}
\tilde{\boldsymbol{\upsigma}}=(\sigma_{xx},\sigma_{yy},\sigma_{zz},\sqrt2\sigma_{yz},
\sqrt2\sigma_{xz},\sqrt2\sigma_{xy}),
\end{equation*}
the pair of Eq.~\eqref{eq:elasticpair}
holds with $\mathbf{G}=\tilde{\mathbf{C}}^{1/2}\mathbf{E}/\sqrt\rho$,
\begin{equation}
  \mathbf{E}=\begin{bmatrix}
    \Db_x & 0 & 0\\ 0 & \Db_y & 0\\ 0 & 0 & \Db_z\\
    0 & \Df_z/\sqrt2 & \Df_y/\sqrt2\\
    \Df_z/\sqrt2 & 0 & \Df_x/\sqrt2\\
    \Df_y/\sqrt2 & \Df_x/\sqrt2 & 0
  \end{bmatrix},
\end{equation}
and $\tilde{\mathbf{C}}^{1/2}$ from Eq.~\eqref{eq:csqrt} with $k=3$:
the normal block has diagonal $(s_1+2s_2)/3$ and off-diagonal
$(s_1-s_2)/3$ with $s_1=\sqrt{3\lambda+2\mu}$, $s_2=\sqrt{2\mu}$, and the
shear block is $s_2\mathbf{I}_3$.  Each normal row of $\tilde{\mathbf{C}}^{1/2}$ has squared
norm $[(s_1+2s_2)^{2}+2(s_1-s_2)^{2}]/9=(s_1^{2}+2s_2^{2})/3
=\lambda+2\mu=\rho c_p^{2}$, as in two dimensions.  The positions of the
staggered grid~\cite{graves1996} are $\tilde v_x$ at
$(i{+}\tfrac12,j,k)$, $\tilde v_y$ at $(i,j{+}\tfrac12,k)$, $\tilde v_z$ at
$(i,j,k{+}\tfrac12)$, normal stresses at nodes, $s_{yz}$ at
$(i,j{+}\tfrac12,k{+}\tfrac12)$, $s_{xz}$ at $(i{+}\tfrac12,j,k{+}\tfrac12)$
and $s_{xy}$ at $(i{+}\tfrac12,j{+}\tfrac12,k)$.  The nine fields occupy
sixteen blocks (four label qubits).  The couplings are nine normal
(each normal stress to each velocity through the full normal block of
$\tilde{\mathbf{C}}^{1/2}$, along that velocity's direction) and six shear; per
direction this is $3+2=5$, so $\mathbf{H}$ has $10\sum_a(n_a+1)$ strings.  The
sponge damps the nine physical blocks with
$\sigma_x+\sigma_y+\sigma_z$, and Theorem~\ref{thm:structured} applies
verbatim with three profiles per block.

%% file: sections/appB.tex
\section{Operator strings and boundary conditions}
\label{app:strings}

For $j=\sum_c2^{c}j_c$ on $n$ qubits, adding one to $j$
flips the trailing ones to zeros and the first zero to one.  The inverse
map $\ket{j+1}\to\ket j$, restricted to the inputs whose lowest $k-1$
bits are $0$ and $k$th bit is $1$, is
$I^{\otimes(n-k)}\otimes\sigma_{01}\otimes\sigma_{10}^{\otimes(k-1)}$.
These $n$ cases partition all $j+1\in\{1,\dots,N-1\}$, which gives
Eq.~\eqref{eq:shift}.  The input $j+1=0$ (i.e.\ the wrap from $N-1$) is
not covered, which is the non-periodic truncation.  Hence
\begin{equation}
  \Df=\frac1h\Bigl(\sum_{k=1}^{n}I^{\otimes(n-k)}\otimes\sigma_{01}\otimes
  \sigma_{10}^{\otimes(k-1)}-\mathbf{I}^{\otimes n}\Bigr),
\end{equation}
and $\Db=-(\Df)^{\dagger}=(\mathbf{I}-\mathbf{S}^{\dagger})/h$ is obtained by
$\sigma_{01}\leftrightarrow\sigma_{10}$ in the shift strings and a sign
change of both the shift and the identity parts.

$(\Df\mathbf{u})_{N-1}=-u_{N-1}/h$ (ghost $u_N=0$) and
$(\Db\mathbf{u})_0=u_0/h$ (ghost $u_{-1}=0$).  For the acoustic system,
$\partial_t^{2}\mathbf{u}=c^{2}\Df\Db\mathbf{u}$ has first row
$c^{2}(-2u_0+u_1)/h^{2}$ (Dirichlet ghost $u_{-1}=0$) and last row
$c^{2}(u_{N-2}-u_{N-1})/h^{2}$ (zero flux at $x_{N-1/2}$, Neumann).  In
terms of $\boldsymbol{\uppsi}$: $q$ vanishes beyond the right edge and $v$ beyond the
left edge.  Both ends reflect totally.  The sponge layers sit inside
these ends, and the reflection from the edge behind the layer is part of
the measured reflection.  For the elastic
system the same truncation applies to every difference in
Eq.~\eqref{eq:stagpair}; the outer edges are again perfect reflectors.

A block coupling
$\ket b\!\bra{b'}\otimes c\,\mathbf{D}_a$ with $\mathbf{D}_a$ a difference on axis $a$
contributes the $n_a+1$ strings of $\mathbf{D}_a$, each prefixed by the label
string of $\ket b\!\bra{b'}$, and its Hermitian partner contributes
another $n_a+1$.  Couplings that share label and axis merge, which is
why the elastic count depends only on the number of couplings per
direction, Eq.~\eqref{eq:termcount}.  We verified numerically that the 2D elastic Hamiltonian has
exactly $6n_x+6n_y+12$ strings for all $1\le n_x,n_y\le5$, and that its matrix
equals the block construction to machine
precision.

%% file: sections/appC.tex
\section{Structured damping circuit: construction details}
\label{app:structured}

The construction has three steps.  First, for each block $\beta$ the damping diagonal
$\Sigma_\beta(i_x,i_y)$ is split as $f_x(i_x)+f_y(i_y)$ by reading off
the row and column through the grid centre, where $\sigma=0$, and the
split is verified exactly.  Blocks with identical profiles on an axis
are grouped, and each group is covered by aligned dyadic ranges of block
labels (sub-cubes of the label register): a range fixes the high label
bits and leaves the low ones free.  For the 2D acoustic sponge the
groups are $\{v,q_y\}$ and $\{q_x\}$ for the $x$ profile (nodal and
staggered) and $\{v,q_x\}$ and $\{q_y\}$ for $y$.  For the elastic
sponge the five physical blocks share both profiles and are covered by
the sub-cube $\{0,1,2,3\}$ and the single label $4$.

Second, a profile on $2^{n_a}$ nodes is
decomposed recursively: on a dyadic interval of size $2^{q}$, compute
the coefficients $a_S$, $|S|\le m$, of Eq.~\eqref{eq:mobius} from the
values at the $\sum_{r\le m}\binom qr$ corner indices.  If the resulting
multilinear polynomial reproduces all $2^{q}$ values, emit the piece;
otherwise bisect.  Intervals on which the profile vanishes are skipped.
The recursion terminates at single nodes at the latest (constants), so
the decomposition is exact for any profile.  For the layers used here
($\nL$ a power of two, $m=2$) each profile has one or two pieces per
layer, plus single-node corrections from clipping and staggering (on the
$8\times8$ grids with $\nL=1$ a whole profile is a single piece).

As a worked example, take $N=16$ nodes, $\nL=4$, $m=2$, nodal sampling:
$\sigma=\sigma_{\max}(4-j)^{2}/16$ on $j<4$,
$\sigma_{\max}(j-11)^{2}/16$ on $j\ge12$, and zero in between.  The left
piece fixes $j_3=j_2=0$ and has $(4-j)^{2}=16-7j_0-12j_1+4j_0j_1$.  The
right piece fixes $j_3=j_2=1$ and, with $j-11=1+j_0+2j_1$ on it,
$(j-11)^{2}=1+3j_0+8j_1+4j_0j_1$.  With
$n_p$ $p$ qubits each of the $2\times4$ monomials becomes $n_p$ phase gates,
e.g.\ the term $4j_0j_1$ of the left piece on $p$ bit $b$ is a phase
$\tau(\pi/p_{\max})\chi_b\,4\sigma_{\max}/16$ on $k_b$, controlled by
$j_0=j_1=1$, $j_2=j_3=0$ and the block-label bits.

Third, each term is one
$\mathrm{C}^{r}P(\theta)$ with target $k_b$.  In the flag variant, the
predicate ``label in sub-cube and high bits of the piece fixed'' is
computed into a clean ancilla by one multi-controlled NOT
with mixed $0/1$ controls, all phases of the piece are controlled on the
ancilla and the monomial bits only, and the predicate is uncomputed.
The ancilla is returned to $\ket0$ exactly; we measure zero leakage in
all runs.

With Qiskit~1.4.5, two properties of the transpiler matter for the flag
variant.  First, by default it assumes that every qubit starts in
$\ket0$ and uses qubits that are still idle as clean ancillas when it
synthesizes multi-controlled gates.  A damping factor transpiled on its
own is then not the same unitary, although every distinct gate of the
circuit compiles exactly on its own and the uncompiled circuit is exact
($10^{-15}$); with this assumption switched off
(\texttt{qubits\_initially\_zero=False}) the whole damping factor
compiles exactly.  Second, for the complete Schr\"odingerized circuit
some optimization levels return the final state with a different global
phase, which is physically irrelevant but enters a direct comparison of
statevectors.  We therefore (i)~simulate the flag variant with Aer at
optimization level~0, where the multi-controlled gates are executed
natively, and verify it against the dense-diagonal pipeline, and (ii)~report its
CNOT counts as the sum of per-gate syntheses, each checked against the
exact gate.  The ancilla-free variant compiles exactly as a whole.